\documentclass[12pt]{article}
\usepackage{amsmath,amssymb,amsfonts}
\usepackage{times}
\usepackage{graphicx}
\usepackage{color}
\usepackage{multirow}
\usepackage[authoryear]{natbib}
\usepackage{rotating}
\usepackage{bbm}
\usepackage{latexsym}
\usepackage{geometry}
\usepackage{multirow}
\usepackage{booktabs} 
\usepackage{array}

\numberwithin{equation}{section}

\begin{document}

\begin{center}
{\LARGE Switching Models of Oscillatory Networks Greatly Improve Inference of Dynamic Functional Connectivity}\end{center}

\ \\
{\bf \large Wan-Chi Hsin$^{\displaystyle 1}$}, {\bf \large Uri T. Eden$^{\displaystyle 1}$} and {\bf \large Emily P. Stephen$^{\displaystyle 1}$}\\
{$^{\displaystyle 1}$Department of Mathematics and Statistics, Boston University, Boston, MA 02215, U.S.A.}\\


\thispagestyle{empty}
\markboth{}{NC instructions}

\begin{center} {\bf Abstract} \end{center}
Functional brain networks can change rapidly as a function of stimuli or cognitive shifts. Tracking dynamic functional connectivity is particularly challenging as it requires estimating the structure of the network at each moment as well as how it is shifting through time. In this paper, we describe a general modeling framework and a set of specific models that provides substantially increased statistical power for estimating rhythmic dynamic networks, based on the assumption that for a particular experiment or task, the network state at any moment is chosen from a discrete set of possible network modes. Each model is comprised of three components: (1) a set of latent switching states that represent transitions between the expression of each network mode; (2) a set of latent oscillators, each characterized by an estimated mean oscillation frequency and an instantaneous phase and amplitude at each time point; and (3) an observation model that relates the observed activity at each electrode to a linear combination of the latent oscillators. We develop an expectation-maximization procedure to estimate the network structure for each switching state and the probability of each state being expressed at each moment. We conduct a set of simulation studies to illustrate the application of these models and quantify their statistical power, even in the face of model misspecification.

\section{Introduction}

Neural oscillations are critical for coordinating information across neurons and brain areas \citep{Singer1999, Fries2001, Fries2005, Buzsaki2012}. When groups of neurons engage in rhythmic activity, they can become functionally connected across large brain distances. This phenomenon can be viewed through the lens of coupled oscillators, where dynamics of neural synchronization reflect changes in information processing and cognitive function \citep{Horwitz2003, Rogers2007}. Changes in brain-wide oscillatory coupling networks can be used to track changes in healthy and diseased brain states and enhance our understanding of the role played by neuronal oscillations \citep{SchnitzlerGross2005, Basar2013, Cocchi2017, Wang2018}. To analyze complex rhythmic interactions in the brain, researchers often rely on descriptive statistics, in particular, spectral estimators, such as coherence and cross-spectral estimates \citep{BuzsakiDraguhn2004, Fries2005}. More advanced statistical modeling methods, such as global coherence analysis, have been developed to understand complex brain activity and functional networks \citep{Mitra2007, Cimenser2011, Wong2011}. However, these descriptive methodologies are difficult to incorporate into statistical models of brain dynamics and are typically restricted to single narrow-band rhythms. More importantly, these approaches are limited in statistical efficiency and temporal resolution in estimating dynamic functional connectivity.

Recently, state-space models have been successful in characterizing rhythmic neural activity in the time-domain \citep{MatsudaKomaki2017, Beck2018, Soulat2022}. State-space models provide a powerful analytic framework across diverse disciplines including control engineering, signal processing, and neuroscience. They are typically characterized by two key equations: the state equation, which describes how the latent state of the system evolves over time, and the observation equation, which relates the observed data to this latent state \citep{ChenBrown2013}. State-space models are particularly effective in analyzing neural signal dynamics because of their generality and ease of interpretation \citep{Kass2014}, and have been widely used in neuroscience to analyze neural dynamics \citep{SmithBrown2003, Eden2004, Chen2010, Paninski2010, Akram2016, Eden2018, Wodeyar2021}. Many of these analyses assume that the statistical properties of the models are time-invariant, but neural activity can shift rapidly between different states such as sleep and wakefulness \citep{Steriade1993}. In this article, we introduce a switching state-space modeling framework that consists of multiple linear state-space models and transitions between these models through time \citep{Murphy1998}. 

The switching Kalman filter, with its capability to capture dynamic transitions by having a bank of different linear models and switching between them, has been applied in various fields. In \citep{Veeraraghavan2005}, it was applied to track vehicles and detect events at traffic intersections, with the goal of improving traffic management systems. In meteorology, it was used to monitor and forecast storm locations in meteorological networks by dynamically adjusting to changing weather conditions \citep{Manfredi2005}. It has also been used to predict potential failures for mechanical systems and proven to be a promising tool to support maintenance decision-making in aerospace, manufacturing, and transportation \citep{LimMba2015}.

Our work is motivated by the challenges associated with conventional windowed methods of estimating dynamic functional connectivity, which have limited statistical power. Time-frequency analyses using sliding windows have been widely applied by the neuroimaging and electrophysiology community to understand how brain dynamics relate to our cognitive behavior \citep{Kucyi2013, KucyiDavis2014, EltonGao2015, ONeill2015, ChangGlover2010, Preti2017}. Consider a scenario where multiple neural sources are described by coupled oscillators. Researchers often compute coherograms, which measure the coherence between pairs of signals as a function of time and frequency, to assess the degree of linear correlation between these signal pairs in the frequency domain. However, the statistical power of coherograms is limited because it relies on relatively small temporal windows for analysis. This constraint impedes our ability to identify connections with confidence over time and, consequently, prevents us from accurately inferring the dynamics of the network structure. 

Rhythmic networks in brain activity may reflect several different mechanisms of coupling: for example, networks may be driven by a common source influencing rhythms in multiple areas, or rhythms may propagate through local networks either by driving each other or through a shared source of noise. In any case, we posit that specific cognitive states and behavioral tasks may be mediated by a small number of network modes whose expression changes dynamically. For instance, during general anesthesia, as a subject transitions from a conscious to unconscious state and back, this change in state is reflected in stereotypical patterns of oscillatory coherence across brain areas \citep{Cimenser2011, Purdon2013}. In our previous work \citep{Hsin2022}, we modeled the first mechanism using a “Common Oscillator Model (COM),” in which the activity across multiple electrodes are linked through a small number of latent oscillators, with each oscillator describing the associated activity of a single network mode. We used an EM algorithm to estimate the composition of each network mode, with a switching Kalman filter and smoother to estimate the instantaneous state of the oscillations driving each network mode. In contrast to windowed methods, such as the coherogram, which makes no assumptions about which links are present at any given time, our approach assumes a discrete set of networks and by switching between them, greatly improves the inference of dynamic functional connectivity.
 
In this work, we first introduce a broad framework for analyzing dynamically coupled oscillators in the brain. In particular, we describe three specific modeling structures that reflect common types of coupling. The fundamental problem we aim to address is the changing functional connectivity among observations, characterized by a small number of distinct functional networks. This modeling framework enables us to estimate both the structures of these functional networks and the times at which switches occur. In addition to the Common Oscillator Model, we develop two other models to capture propagating rhythms. The first we call the Correlated Noise Model (CNM), which defines networks linked by correlation structure in the noise driving the oscillations on each electrode. The second we call the Directed Influence Model (DIM), which allows the instantaneous state of one oscillating source to directly influence another. All these methods use the full data to estimate a discrete set of functional network modes and use local data over short time scales to estimate which network modes are being expressed at each moment. Each model is comprised of three components: (1) a set of latent switching states that represent transitions between the expression of each network mode; (2) a set of latent oscillators, each characterized by an estimated mean oscillation frequency and an instantaneous phase and amplitude at each time point; and (3) an observation model that relates the observed activity at each electrode to a linear combination of the latent oscillators. We use switching Kalman filters and smoothers to estimate the instantaneous phase and amplitude of each oscillator and the probability of each switching state at every time step. The estimated switching state characterizes which of the network modes is being expressed at each time point. 



\section{Methods}

\noindent We begin by introducing a broad analytic framework that uses latent processes to represent the oscillators that drive brain dynamics, and subsequently derive three specific model structures to show different ways that network structure can be captured. In our approach, we extend the oscillation decomposition framework described in \citep{MatsudaKomaki2017} and use state models representing sets of oscillators. Each oscillator comprises a pair of latent variables with coupled dynamics. For example, we model a single oscillator as a random vector process $x_t = [x_{t,\text{Re}}, x_{t,\text{Im}}]'$, with one-step update equation $x_{t} = Ax_{t-1}+\epsilon_t$, where $A$ is a $2 \times 2$ rotation matrix of the form $A = a \begin{pmatrix} \text{cos}(\omega) & -\text{sin}(\omega)\\ \text{sin}(\omega) & \text{cos}(\omega) \end{pmatrix}$, $a$ is an autoregressive parameter, and $\omega$ is the rotation frequency. Note that we write $x_{t,\text{Re}}$ and $x_{t,\text{Im}}$ to evoke the real and imaginary parts of the analytic representation of a narrow-band signal, $z(t)=x(t)+ix_H(t)$, where $H$ represents the Hilbert transform (See \citep{MatsudaKomaki2017} for more details), even though all random variables are real valued.

More generally, let $K$ denote the number of oscillators. The $2K \times 1$ vector $x_t = [x^1_{t,\text{Re}},x^1_{t,\text{Im}} \ldots, x^K_{t,\text{Re}},x^K_{t,\text{Im}}]'$ is defined as the hidden state variable with $K$ oscillators at time $t$, where $[x^k_{t,\text{Re}},x^k_{t,\text{Im}}]'$ is the state of  oscillator $k$ at time $t$. 

Next, we define a discrete switching state $S_t$ which takes on one of a small number of discrete states at each time step. The switching state determines the dynamics of the oscillatory state $x_t$ at each time step and the influence of the state on the observed data. Importantly, $S_t$ will control which functional network is expressed at time $t$. We assume that $S_t$ is a Markov chain, and denote its transition matrix $Z$, whose $(i,j)^{th}$ element defines the probability of a switch between time steps $t-1$ and $t$.  
\begin{equation}
	P(S_t=j|S_{t-1}=i) = Z_{ij}. 
\end{equation}

The oscillatory latent process model is defined so as to depend on the current value of the switching state: 
\begin{equation}
(x_t|x_{t-1},S_t=j) \sim N(A_j x_{t-1},\Sigma_j),\label{global-state}
\end{equation}
where $A_j$ is a $2K \times 2K$ state transition matrix defining the dynamics of all $K$ oscillators in the current switching state, and $\Sigma_j$ is a $2K \times 2K$ process noise covariance matrix.

The data recorded at each of $N$ neural sources is modeled as an $N$-dimensional multivariate time series, $y_t = [y_t^1,\ldots,y_t^N]'$. The observation equation is also defined so as to depend on the current value of the switching state:
\begin{equation}
	(y_t|x_t,S_t=j) \sim N(B_j x_t,R),\label{global-obs}
\end{equation}
where $B_j$ is an $N \times 2K$ observation matrix defining the influence of each of the two components of each of the $K$ oscillators on each of the $N$ neural sources, and $R$ is an $N \times N$ observation noise covariance matrix.

There are multiple ways that the $N$ neural sources can be coupled in this general model framework. In particular, options include: (1) using the observation matrices, $B_j$, to express how the activity of a small number of latent oscillators are reflected across many neural sources, (2) using the state covariance matrices, $\Sigma_j$, to express shared noise structure propagating through the network, or (3) using the state update matrices, $A_j$, to express the direct influence of one source on each of the others. In the following three subsections, we derive specific models to capture functional networks based on these mechanisms.

\subsection{Common Oscillator Model}\label{com} 

\begin{figure}[htbp]
\centerline{\includegraphics[scale=0.5]{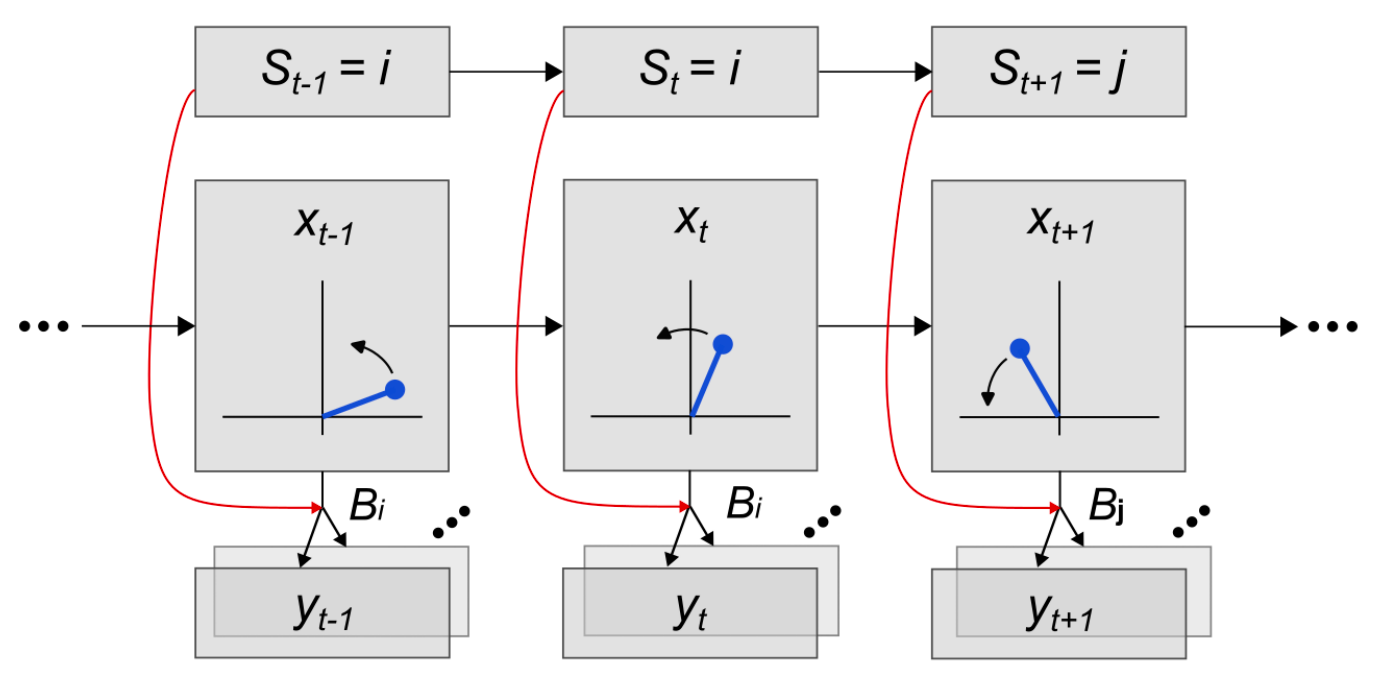}}
\caption{Schematic representation of the Common Oscillator Model. The observations $y_t$ are a linear function of the continuous latent oscillators $x_t$, and the observation matrix $B$ depends on the discrete latent state $S_t$. This schematic shows an example with one oscillator driving two neural sources.}
\label{fig-COM}
\end{figure}

For this model, we posit that neural sources are linked by being driven by a common set of oscillators. The network structure is reflected in the observation matrix, which determines which of the oscillators is driving each node and the relation between the instantaneous phase and amplitudes of the oscillators and each observed signal. A schematic for this model is shown in Figure \ref{fig-COM}, where the red arrows going from switching state variable $S_t$ to the observations $y_t$ indicate that the observation matrix $B_j$ changes over time depending on the switching state. Note that in this model, the state transition matrix $A$ and the state noise covariance $\Sigma$ do not depend on the switching state $S_t$, so are assumed to be constant in time. When $B$ switches, the functional network driven by each oscillator changes as well. This model structure is particularly efficient when there are a small number of latent oscillators $K$ relative to the number of sources $N$, so we show the arrow from $x_t$ to $y_t$ splitting. This schematic displays an example of one oscillator driving a pair of observed signals. 

The latent process model is defined as 
\begin{equation}
(x_t|x_{t-1}) \sim N(A x_{t-1},\Sigma),\label{COM-state}
\end{equation}
where
$$A = \begin{pmatrix}
A_1 & 0 & \cdots & 0 \\
0 & A_2 & \cdots & 0 \\
\vdots & \vdots & \ddots & \vdots \\
0 & 0 & \cdots & A_K 
	\end{pmatrix}, 
$$
and 
$$\Sigma = \begin{pmatrix} 
\sigma_1 I_{2\times2} & 0 & \cdots & 0 \\
0 & \sigma_2 I_{2\times2} & \cdots & 0 \\
\vdots & \vdots & \ddots & \vdots \\
0 & 0 & \cdots & \sigma_K I_{2\times2}
	\end{pmatrix}.$$

\noindent $A$ is a $2K \times 2K$ block diagonal matrix composed of $A_1,\ldots,A_K$. For $k=1,\ldots,K$, $$A_k = a_k \begin{pmatrix}
\text{cos}(\omega_k) & -\text{sin}(\omega_k) \\
\text{sin}(\omega_k) & \text{cos}(\omega_k)
	\end{pmatrix}, $$ and $\omega_k = 2\pi f_k/F_s,$ where $f_k$ is the oscillation frequency and $F_s$ is the sampling frequency.
Each $A_k$ is a 2-dimensional rotation matrix multiplied by an autoregressive parameter $a_k \in (0,1)$. $\Sigma$ is a $2K \times 2K$ diagonal noise covariance matrix. 

For an $N$-dimensional multivariate time series sampled at frequency $F_s$, the observation at time $t$ is $y_t = [y_t^1,\ldots,y_t^N]'$, where $y_t^n$ is the data recorded at node $n$ at time $t$. The observation equation is given by
\begin{equation}
	(y_t|x_t,S_t=j) \sim N(B_j x_t,R).\label{COM-obs}
\end{equation}

The observation process $y_t$ is the weighted sum of contributions from one or more oscillatory latent states. These contributions are determined by the $N\times 2K$ observation matrix $B_j$, where the subscript $j$ indicates the matrix is associated with switching state $S_t = j$. Each switching state has its own observation matrix, and each of these matrices represents a distinct network mode. The observation noise is Gaussian with $N\times N$ covariance matrix $R$. We can write the observation matrix for state $j$ elementwise: 

\begin{equation}\label{B-eq}
  B_j = \begin{pmatrix}
	b_{j,\text{Re}}^{1,1} & b_{j,\text{Im}}^{1,1} & \cdots & b_{j,\text{Re}}^{1,K} & b_{j,\text{Im}}^{1,K} \\
	\vdots & \vdots & \ddots & \vdots & \vdots\\
 b_{j,\text{Re}}^{N,1} & b_{j,\text{Im}}^{N,1} & \cdots & b_{j,\text{Re}}^{N,K} & b_{j,\text{Im}}^{N,K} \end{pmatrix}	
\end{equation}

The elements $b_{j,\text{Re}}^{n,k}$ and $b_{j,\text{Im}}^{n,k}$ represent the respective influence of the first and second components of the $k^{\text{th}}$ oscillator on the $n^{\text{th}}$ node. The amplitude $\sqrt{(b_{j,\text{Re}}^{n,k})^2 + (b_{j,\text{Im}}^{n,k})^2}$ describes the degree to which the oscillator $k$ that is expressed at node $n$ in state $j$. For any two nodes, $n_1$ and $n_2$ driven by the same oscillator $k$, the phase difference between them is equal to $\theta_{n_1,n_2} = \theta_{n_1} - \theta_{n_2}$, where $\theta_{n_i}$ is the argument of the complex number $b_{j,\text{Re}}^{n_i,k} + i b_{j,\text{Im}}^{n_i,k}$.

By comparing the state equation \ref{COM-state} and the observation model \ref{COM-obs} of the Common Oscillator Model with those from the global model, equations \ref{global-state} and \ref{global-obs}, the key difference is that we suppress the the subscript $j$ in the state transition matrix $A$ and the process noise covariance $\Sigma$. In the COM, we use the observation model to express the network structure. This is represented by the observation matrix $B_j$, which changes over time depending on the switching state variable $S_t$.

\subsection{Correlated Noise Model}\label{cnm}

\begin{figure}[htbp]
\centerline{\includegraphics[scale=0.42]{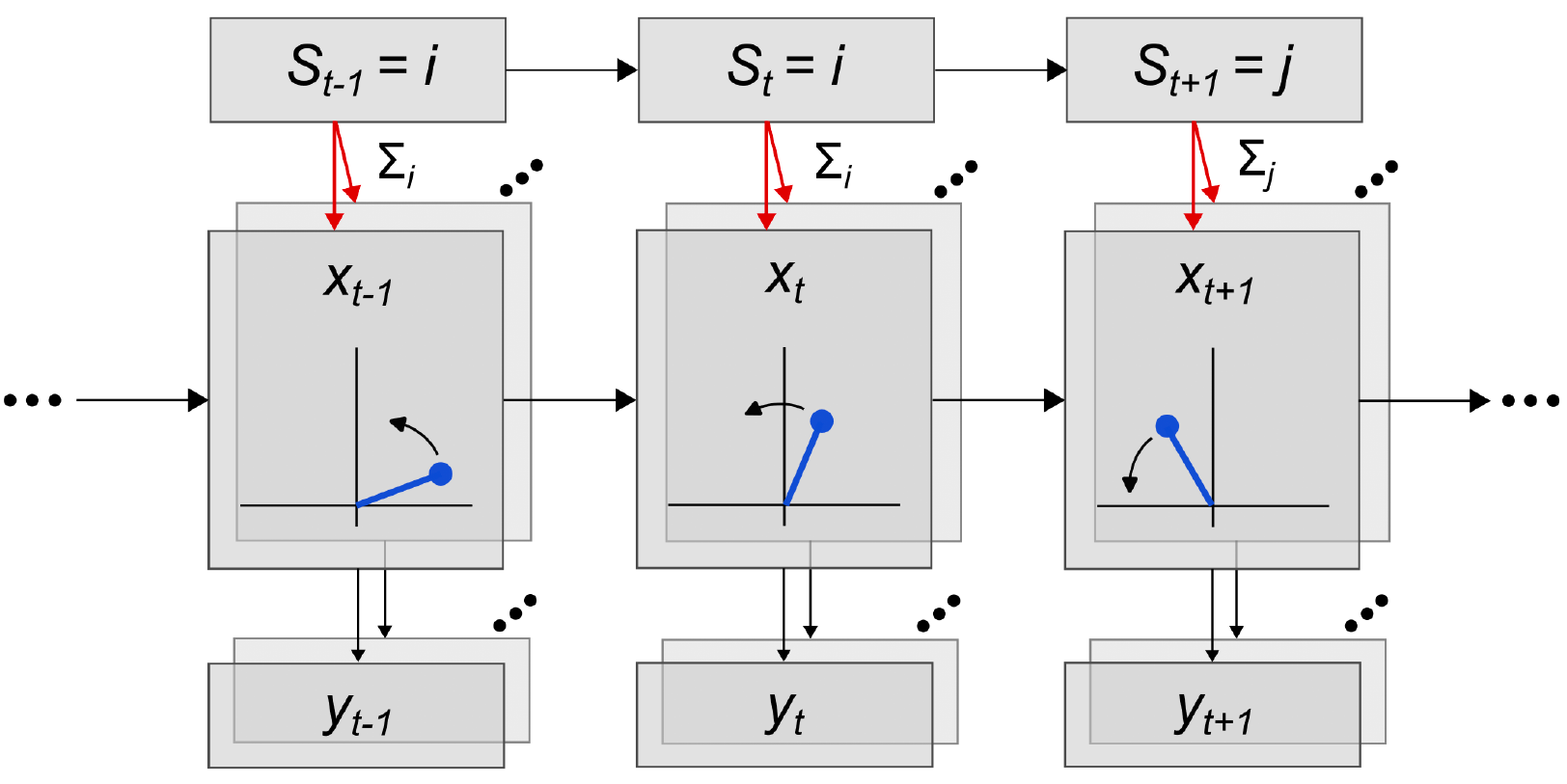}}
\caption{Schematic representation of the Correlated Noise Model. The latent oscillators $x_t$ is directly influenced by the discrete latent state $S_t$ through the process noise covariance matrix $\Sigma$.}
\label{fig-CNM}
\end{figure}

In contrast to the COM, the Correlated Noise Model (CNM) assumes that each node is characterized by its own oscillator, any combination of which can be linked through correlated noise structure. The network structure is then reflected in the process noise covariance matrix. In Figure \ref{fig-CNM}, the red arrows from the switching state variable $S_t$ to the latent oscillatory state $x_t$ indicate that the process noise covariance $\Sigma_j$ changes through time, depending on the switching state $S_t$. In this model, the state transition matrix $A$ and the observation matrix $B$ do not depend on the switching state $S_t$. Note that the number of layers in $y_t$ is same as the number of layers in $x_t$, since the number of observations is equal to the number of oscillators, with one oscillator per node.

The latent process model is defined as 
\begin{equation}
(x_t|x_{t-1}, S_t=j) \sim N(A x_{t-1},\Sigma_j),\label{eq2}
\end{equation}
where the $A$ matrix shares the same structure as the $A$ matrix described in the Common Oscillator Model in Section \ref{com}.

\noindent The process noise covariance matrix is
 \begin{equation}\label{Q-matrix}
	    \begin{split} & \Sigma_j = \begin{pmatrix}
	        \Sigma_j^1 & \Sigma_j^{1,2} & \cdots & \Sigma_j^{1,K} \\
	        \Sigma_j^{2,1} & \Sigma_j^2 & \cdots & \Sigma_j^{2,K} \\
	        \vdots & \vdots & \ddots & \vdots \\
	        \Sigma_j^{K,1} & \Sigma_j^{K,2} & \cdots & \Sigma_j^K
	    \end{pmatrix}, \text{where} \\
        \\
        & \Sigma_j^{k} = \sigma_j^{k}I_{2 \times 2}, \text{and} \\
	    & \Sigma_j^{n_1,n_2} = \left\{ \begin{array}{ll} \rho_j^{n_1,n_2} \begin{pmatrix} \mbox{cos}(\theta_j^{n_1,n_2}) & -\mbox{sin}(\theta_j^{n_1,n_2}) \\ \mbox{sin}(\theta_j^{n_1,n_2}) & \mbox{cos}(\theta_j^{n_1,n_2}) & \end{pmatrix}, & \text{if $n_1$ and $n_2$ are linked.} \\
	    0_{2\times2}, & \text{otherwise.}
	    \end{array}\right.
	    \end{split}    
    \end{equation}

\noindent The amplitude $\rho_j^{n_1,n_2}$ describes the degree to which the nodes $n_1$ and $n_2$ are linked and $\theta_j^{n_1,n_2}$ is the phase difference between these two nodes in the switching state $j$. The rotational structure in the off-diagonal blocks indicates that when one node receives a random perturbation, the other node to which it is linked receives a related random perturbation in a phase-dependent manner. Note that the process noise covariance $\Sigma_j$ is symmetric. For example, if $n_1$ and $n_2$ are linked with a 90-degree phase difference in state $j$ ($\theta_j^{n_1,n_2} = \pi/2$), then $n_2$ and $n_1$ are linked with a 270-degree (-90-degree) phase difference ($\theta_j^{n_2,n_1} = -\pi/2$). 

Under this model, the observation equation is given by
\begin{equation}
	(y_t|x_t) \sim N(B x_t,R), 
\end{equation}
where $B$ is a $K \times 2K$ block diagonal matrix
$$ B = \begin{pmatrix}
	B_1 & 0 & \cdots & 0 \\
    0 & B_2 & \cdots & 0 \\
	\vdots & \vdots & \ddots & \vdots\\
	0 & 0 & \cdots & B_K
\end{pmatrix}	$$
and $B_k = [b^k_{\text{Re}}, b^k_{\text{Im}}]$ determines the extent to which the real and imaginary components of the oscillator are expressed in the data. For our simulations we opted to set $b^k_{\text{Re}} = 1$ and $b^k_{\text{Im}} = 0$ so that the signal at each node is determined by only the real component of the oscillator. 

The main difference between the global model descried in equations \ref{global-state} and \ref{global-obs} and the specific CNM structure, is the suppression the subscript $j$ in the state transition matrix $A$ and the observation matrix $B$. In the CNM, the state noise covariance $\Sigma_j$ determines the network structure, and it changes over time depending on the switching state variable $S_t$.

\subsection{Directed Influence Model}\label{dim}

\begin{figure}[htbp]
\centerline{\includegraphics[scale=0.42]{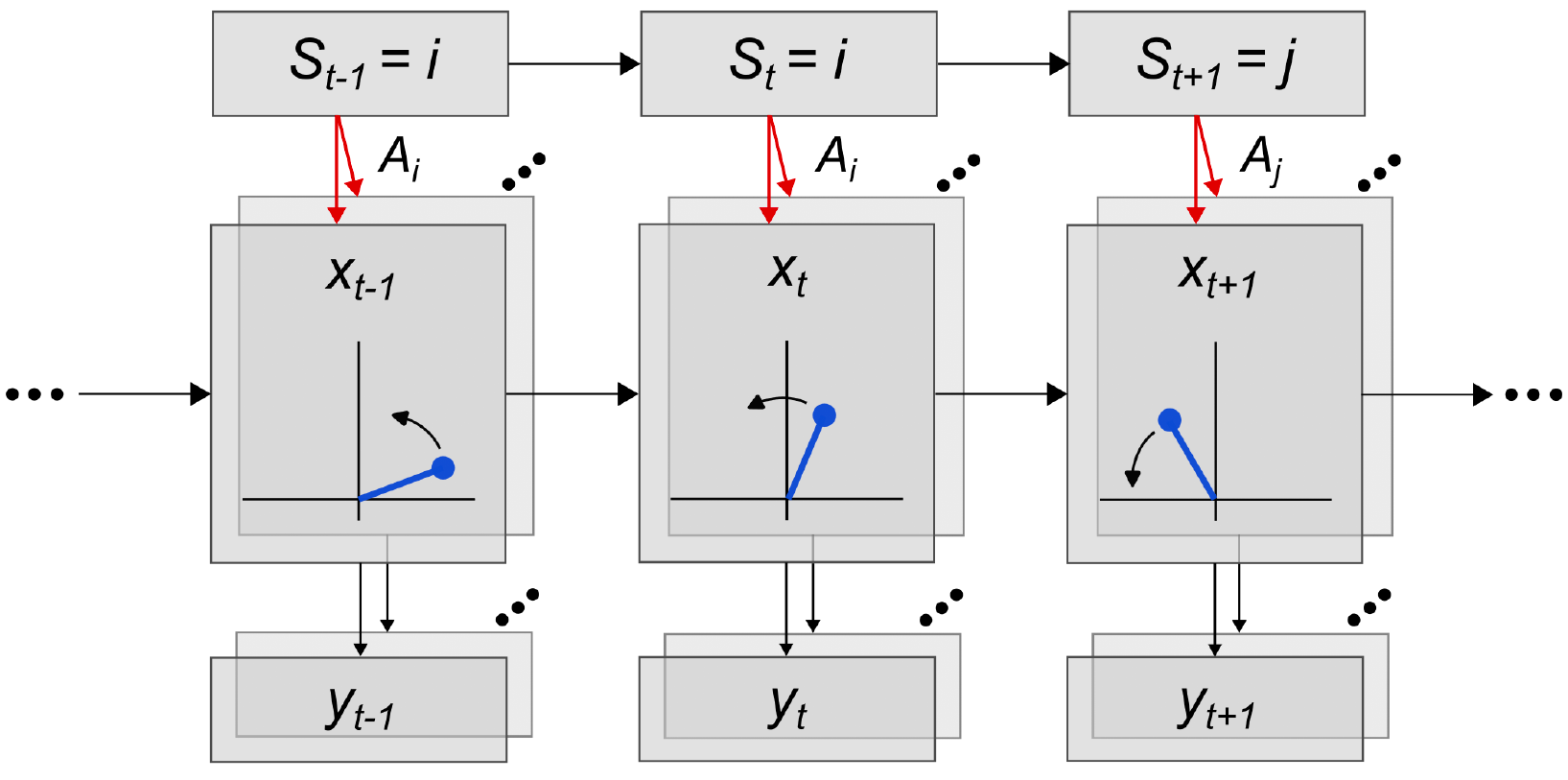}}
\caption{Schematic representation of the Directed Influence Model. The latent oscillators $x_t$ are directly influenced by the discrete latent state $S_t$ through the state transition matrix $A$.}
\label{fig-DIM}
\end{figure}

The last model structure we develop is the Directed Influence Model (DIM), in which each neural source is characterized by its own oscillator, and the state of one oscillator directly influences another. The network structure is then determined by the state transition matrix. In Figure \ref{fig-DIM}, the red arrows going from the switching state variable $S_t$ to the latent oscillatory state $x_t$ illustrates that the switching state variable determines which state transition matrix $A_j$ is used at any given time. In this model, the observation noise covariance $\Sigma$ and the observation matrix $B$ do not depend on the switching state $S_t$. Note that, similar to the CNM, the number of oscillators is same as the number of neural sources.

\noindent The latent process model is defined as 
\begin{equation}
(x_t|x_{t-1}, S_t=j) \sim N(A_j x_{t-1},\Sigma),\label{eq3}
\end{equation}
where the process noise covariance matrix $\Sigma$ has the same structure as in the Common Oscillator Model in Section \ref{com}. The state transition matrix $A_j$ is defined as
\begin{equation} \label{A-matrix}
	    \begin{split} & A_j = \begin{pmatrix}
	        A_j^1 & A_j^{1,2} & \cdots & A_j^{1,K} \\
	        A_j^{2,1} & A_j^2 & \cdots & A_j^{2,K} \\
	        \vdots & \vdots & \ddots & \vdots \\
	        A_j^{K,1} & A_j^{K,2} & \cdots & A_j^K
	    \end{pmatrix}, \text{where} \\
        \\
        & A_j^{k} = \alpha_j^{k} \begin{pmatrix} \mbox{cos}(2\pi f_n/F_s) & -\mbox{sin}(2\pi f_n/F_s) \\ \mbox{sin}(2\pi f_n/F_s) & \mbox{cos}(2\pi f_n/F_s) & \end{pmatrix} - \displaystyle\sum_{n_1 = k} \alpha_j^{n_1,n_2} \begin{pmatrix} 1 & 0 \\ 0 & 1 \end{pmatrix}, \text{and} \\
	    & A_j^{n_1,n_2} = \left\{ \begin{array}{ll} \alpha_j^{n_1,n_2} \begin{pmatrix} \mbox{cos}(\phi_j^{n_1,n_2}) & -\mbox{sin}(\phi_j^{n_1,n_2}) \\ \mbox{sin}(\phi_j^{n_1,n_2}) & \mbox{cos}(\phi_j^{n_1,n_2}) \end{pmatrix}, & \text{if $n_1$ and $n_2$ are linked.} \\
	    0_{2\times2}, & \text{otherwise.}
	    \end{array}\right.
	    \end{split}    
    \end{equation}

\noindent The amplitude $\alpha_j^{n_1,n_2}$ represents the coupling strength between nodes $n_1$ and $n_2$, and $\phi_j^{n_1,n_2}$ is the phase difference between these two nodes in switching state $j$. The diagonal blocks include one term that is a rotation matrix, representing the oscillator for the corresponding node, and a second adjustment term that is a row sum of coupling strengths present on the off-diagonal blocks, multiplied by an identity matrix. This adjustment term comes from the influences received from other nodes. Note that the state transition matrix $A_j$ is not necessarily symmetric.  

As with the CNM, under this model, the observation equation is given by
\begin{equation}
	(y_t|x_t) \sim N(B x_t,R), 
\end{equation}
where $B$ is a $K \times 2K$ block diagonal matrix
$$ B = \begin{pmatrix}
	B_1 & 0 & \cdots & 0 \\
    0 & B_2 & \cdots & 0 \\
	\vdots & \vdots & \ddots & \vdots\\
	0 & 0 & \cdots & B_K
\end{pmatrix}	$$
and $B_k = [b^k_{\text{Re}}, b^k_{\text{Im}}]$ determines the extent to which the real and imaginary components of the oscillator are expressed at each node. For the sake of interpretability, we opted to set $b^k_{\text{Re}} = \frac{1}{\sqrt{2}}$ and $b^k_{\text{Im}} = \frac{1}{\sqrt{2}}$ so that the signal at each node is determined equally by the real and imaginary components of the oscillator. If we were to use an observation matrix that takes only the real parts, it would imply that the imaginary component of each oscillator remains unconstrained across oscillators, allowing for a mixing of oscillators when we estimate the A matrix. 

In the DIM, we remove the subscript $j$ from the process noise covariance $\Sigma$ and the observation matrix $B$ in the global model within equations \ref{global-state} and \ref{global-obs}. Here, the state transition matrix $A_j$ describes how different latent sources influence each other in a directed manner, leading to directed functional connectivity between these nodes. We note that this is the only one of these three models that captures directional influences.

\subsection{Oscillatory and Switching State Estimation}\label{latent_est}

We aim to estimate the latent oscillatory state $x_t$ and the switching state $S_t$ given the observations $y_t$. However, the posterior distribution of $x_t$ is a mixture of Gaussians where the number of mixture components grows exponentially in time so that there would be $M^t$ components by time step $t$. To deal with this exponential growth, we utilize a collapse approach, which uses the Generalized Pseudo Bayesian algorithm \citep{BarShalom2001, LiZhang2000} to approximate the mixture of $M^t$ Gaussians using only $M$ components at each time step $t$. 

We compute the posterior smoothing distributions $P(x_t | y_{1:T})$ and $P(S_t = j | y_{1:T})$, of the latent states up to time $t$, given all of the observations up to time $T$ in two steps. We first apply the Kalman filter algorithm \citep{Kalman1960, Sarkka2013} to compute the filter distributions, $P(x_t | y_{1:t})$ and $P(S_t = j | y_{1:t})$, at each time step. The propagation from time $t-1$ to $t$ is given by 
\begin{equation*}P(x_t | y_{1:t},S_{t-1}=i,S_t=j) = N(x_t^{ij},V_t^{ij}),\label{2b1}
\end{equation*}
where 
\begin{equation*}
\begin{split}
x_t^{ij} & = \text{E}(x_t | y_{1:t},S_{t-1}=i,S_t=j),\\
 V_t^{ij} & = \text{Cov}(x_t | y_{1:t},S_{t-1}=i,S_t=j). 
 \end{split}
\end{equation*}

\noindent Let $L_t^{ij} = P(y_t | y_{1:t-1},S_{t-1}=i,S_t=j)$ be the likelihood of observing $y_t$, and denote
\begin{equation*}
\begin{split}
    w_t^{ij} & = P(S_{t-1}=i,S_t=j | y_{1:t}), \\
    w_t^j & = P(S_t=j | y_{1:t}), \\
    w_t^{i|j} & = P(S_{t-1}=i | S_t=j, y_{1:t}).
 \end{split}
\end{equation*}
We then compute the following weights: 
\begin{equation*}
\begin{split}
    w_t^{ij} & = \frac{L_t^{ij} Z_{ij} w_{t-1}^i}{\sum_{k}\sum_{l} L_t^{kl} Z_{kl} w_{t-1}^k}, \\
    w_t^j & = \sum_i w_t^{ij}, \\
    w_t^{i|j} & = \frac{w_t^{ij}}{w_t^j}.
 \end{split}
\end{equation*}

\noindent The posterior filter distribution is
\begin{equation*}P(x_t | y_{1:t}) = \sum_{i=1}^{M} \sum_{j=1}^{M} w_t^{ij} P(x_t | y_{1:t},S_{t-1}=i,S_t=j),
\end{equation*}
which is a mixture of $M^2$ Gaussians. To collapse the mixture of $M^2$ Gaussians into a mixture of $M$ Gaussians, we apply the collapse operator as described in \citep{Murphy1998} to combine the estimates from the filters in the previous time step with the weight $w_t^{i|j}$. That is, for each $j$,
\begin{equation*}
    P(x_t | y_{1:t},S_t=j) = \sum_{i=1}^{M} w_t^{i|j} P(x_t | y_{1:t},S_{t-1}=i,S_t=j),
\end{equation*}
is a mixture of $M$ components and we approximate this as a single Gaussian by applying the collapse operator detailed in the Appendix, so that we have
$$P(x_t | y_{1:t},S_t=j) \approx N(x_t^{j},V_t^{j}),$$
where 
\begin{equation*}
\begin{split} 
    x_t^{j} & = \text{E}(x_t | y_{1:t},S_t=j), \\
    V_t^{j} & = \text{Cov}(x_t | y_{1:t},S_t=j).
 \end{split}
\end{equation*}
Then, the posterior probability of the hidden state conditioned on the observations can be computed by 
\begin{equation*}
    P(x_t | y_{1:t}) \approx \sum_{j=1}^M w_t^j P(x_t | y_{1:t},S_t=j)
\end{equation*}
where the posterior distribution $P(x_t | y_{1:t})$ is approximated by a mixture of $M$ Gaussians at each time step $t$. Note that if the goal is to compute $P(x_t | y_{1:T})$, we do not have to apply the collapse operator again for $\hat{x_t}$ and $V_t$ where $\hat{x_t} = \text{E}(x_t|y_{1:t})$ and $V_t = \text{Cov}(x_t|y_{1:t})$, instead, we will pass $x_t^j$ and $V_t^j$ on to the smoothing step.

In the smoothing step, we estimate $P(x_t | y_{1:T})$ and $P(S_t = j | y_{1:T})$ by using all of the data including both past and future observations. Similar to the filter step, we have
$$P(x_t | y_{1:T},S_t=j, S_{t+1}=k) = N(\tilde{x}_t^{jk},\tilde{V}_t^{jk}),$$
where 
\begin{equation*}
\begin{split} 
\tilde{x}_t^{jk} & = \text{E}(x_t | y_{1:T},S_t=j, S_{t+1}=k),\\
\tilde{V}_t^{jk}  & = \text{Cov}(x_t | y_{1:T},S_t=j, S_{t+1}=k).  
 \end{split}
\end{equation*}
To obtain $\tilde{x}_t^{jk}$ and $\tilde{V}_t^{jk}$, we perform the Rauch-Tung-Strieber (RTS) smoothing algorithm \citep{Rauch1965, Sarkka2013} starting at the last time step $T$ and stepping backwards to the first time step. We denote
\begin{equation*}
\begin{split} 
    g_t^{jk} & = P(S_t = j, S_{t+1} = k|y_{1:T}),\\
    g_t^j & = P(S_t = j | y_{1:T}), \\
    g_t^{k|j} & = P(S_{t+1} = k | S_t = j, y_{1:T}).
 \end{split}
\end{equation*}
We then compute the following weights:
\begin{equation*}
\begin{split} 
    g_t^{jk} & \approx \frac{w_t^j Z_{jk}}{\sum_{j'} w_t^{j'} Z_{j'k}},\\
    g_t^j & = \sum_k g_t^{jk}, \\
    g_t^{k|j} & = \frac{g_t^{jk}}{g_t^j}.
 \end{split}
\end{equation*}
The approximation of $g_t^{k|j}$ follows the standard approach for dynamic linear models (DLMs) with Markov-switching \citep{Kim1994, Murphy1998}. We use the collapse operator to combine the estimates from the smoothers and the weight $g_t^{k|j}$ so that we have $$P(x_t | y_{1:T},S_t=j) \approx N(\tilde{x}_t^{j},\tilde{V}_t^{j}),$$
where 
\begin{equation*}
\begin{split} 
    \tilde{x}_t^{j} & = \text{E}(x_t | y_{1:T},S_t=j), \\
    \tilde{V}_t^{j} & = \text{Cov}(x_t | y_{1:T},S_t=j). 
 \end{split}
\end{equation*}
Then, the posterior probability of the hidden oscillatory state can be computed by 
$$P(x_t|y_{1:T}) \approx \sum_{j=1}^M g_t^j P(x_t | y_{1:T},S_t=j),$$
where $P(x_t|y_{1:T})$ is approximated by a mixture of $M$ Gaussians at every time step $t$. Finally, to obtain the posterior smoothing distribution $$P(x_t | y_{1:T}) = N(\tilde{x}_t, \tilde{V}_t),$$ where 
\begin{equation*}
\begin{split}
    \tilde{x}_t & = \text{E}(x_t | y_{1:T}), \\
    \tilde{V}_t & = \text{Cov}(x_t | y_{1:T}),
 \end{split}
\end{equation*}
we perform the collapse operator again to obtain a single estimate $\tilde{x}_t$ and its uncertainty $\tilde{V}_t$.

\noindent In summary, we apply the Kalman filter and RTS smoother algorithms to estimate the oscillatory and switching states in sequence as follows. The details of the Filter, Smoother, and Collapse operators are outlined in the Appendix. 

\noindent The forward step for filtering from time $t=1$ to $t=T$:
\begin{align*}
    (x_t^{ij}, V_t^{ij}, L_t^{ij}) & = \text{Filter}(x_{t-1}^i, V_{t-1}^i, y_t, A_j, \Sigma_j, B_j, R)\\
    w_t^{ij} & = \frac{L_t^{ij} Z_{ij} w_{t-1}^i}{\sum_{i}\sum_{j} L_t^{ij} Z_{ij} w_{t-1}^i},\\
    w_t^j & = \sum_i w_t^{ij}, \\
    w_t^{i|j} & = w_t^{ij}/w_t^j,\\
    (x_t^j, V_t^j) & = \text{Collapse}(x_t^{ij},V_t^{ij},w_t^{i|j})\\
\end{align*}
The backward step for smoothing from time $t=T$ to $t=1$:
\begin{align*}
    (\tilde{x}_t^{jk}, \tilde{V}_t^{jk}, \tilde{V}_{t+1,t}^{jk}) & = \text{Smoother}(\tilde{x}_{t+1}^k, \tilde{V}_{t+1}^k, x_t^j, V_t^j, A_j, \Sigma_j)\\
    g_t^{jk} & \approx \frac{w_t^j Z_{jk}}{\sum_{j'} w_t^{j'} Z_{j'k}},\\
    g_t^j & = \sum_k g_t^{jk}, \\
    g_t^{k|j} & = g_t^{jk}/g_t^j,\\
    (\tilde{x}_t^j, \tilde{V}_t^j) & = \text{Collapse}(\tilde{x}_t^{jk},\tilde{V}_t^{jk},g_t^{k|j})\\
    (\tilde{x}_t, \tilde{V}_t) & = \text{Collapse}(\tilde{x}_t^{j},\tilde{V}_t^{j},g_t^{j})
\end{align*}
To obtain the one-step covariance term, $\tilde{V}_{t+1,t} = \text{Cov}(x_{t+1}, x_{t} | y_{1:T})$, we can perform the cross-collapse operator (details in the Appendix) twice on $\tilde{V}_{t+1,t}^{jk} = \text{Cov}(x_{t+1}, x_{t} | y_{1:T}, S_{t}=j, S_{t+1}=k)$, which is computed in the smoother step. Note that depending on which one of these three model structures is applied, $A_j$, $\Sigma_j$, and $B_j$ will be fixed as $A$, $\Sigma$, and $B$, accordingly.

\subsection{Model Parameter Estimation}\label{parameter_est}

We develop an expectation-maximization (EM) algorithm \citep{Dempster1977} to estimate simultaneously the model parameters $\theta = \{A_j, \Sigma_j, B_j, R, Z\}$ and the values of the latent switching state $S_t$ and oscillatory state $x_t$ at each time $t$. The likelihood of the complete data is given by
\begin{equation*}
P(x_{1:T}, S_{1:T}, y_{1:T}) = \prod_{t=1}^T P(y_t | x_t, S_t) \prod_{t=2}^T P(x_t|x_{t-1}, S_t) \prod_{t=2}^T P(S_t | S_{t-1}) P(x_1) P(S_1),
\end{equation*}
where
\begin{equation}
\begin{split}
	& P(x_t|x_{t-1}, S_t=j) = (2\pi)^{-\frac{1}{2}} |\Sigma_j|^{-\frac{1}{2}} \text{exp}\left[ -\frac{1}{2}(x_t - A_jx_{t-1})^{'}\Sigma_j^{-1} (x_t - A_jx_{t-1}) \right], \\
    & P(y_t | x_t, S_t=j) = (2\pi)^{-\frac{1}{2}} |R|^{-\frac{1}{2}} \text{exp}\left[ -\frac{1}{2}(y_t - B_jx_{t})^{'} R^{-1} (y_t - B_j x_{t}) \right],\\
	& P(S_t=j | S_{t-1}=i) = Z_{ij}, \\
    & P(x_1) = (2\pi)^{-\frac{1}{2}}|C_1|^{-\frac{1}{2}}\text{exp}\left[ -\frac{1}{2}(x_1 - \mu_1)^{'} C_1^{-1} (x_1 - \mu_1) \right].
\end{split}
\end{equation}
Note that $\mu_1$ and $C_1$ are the initial mean and covariance of the initial latent oscillatory state $x_1$.

In the EM algorithm, we iteratively maximize the expected value of the complete data log likelihood with respect to $\theta$. The E step involves computing the expected value of the complete data log likelihood function of $\theta$, which is also called the $Q$ function:
 \begin{equation}
 \begin{split}
 Q(\theta) & = \text{E}[\text{log}(P(x_{1:T}, S_{1:T}, y_{1:T}))] \\ 
 & \circeq -\frac{1}{2}  \sum_{t=2}^T \sum_{S_t=j} g_t^j \left[ \text{E} \left[ (x_t-A_j x_{t-1})'\Sigma_j^{-1}(x_t-A_j x_{t-1}) - \text{log}|\Sigma_j| \right]\right]\\
 & \quad -\frac{1}{2}  \sum_{t=1}^T \sum_{S_t=j} g_t^j \left[ \text{E} \left[(y_t-B_j x_{t})'R^{-1}(y_t-B_j x_{t})\right]\right] -\frac{T}{2} \text{log}|R| ,
 \end{split}
 \end{equation}
where $g_t^j = P(S_t = j | y_{1:T})$. The details of the derivation process of the E-step of the switching Kalman filter are outlined in \citep{Murphy1998}. The M-step involves maximizing the $Q$ function by taking its derivative with respect to $\theta$ and setting it to 0. The closed-form solutions for $A_j$, $\Sigma_j$, and $B_j$ can be found in \citep{Murphy1998}, which are essentially weighted versions of the standard closed-form solutions of the linear Gaussian state-space models in \citep{Ghahramani1996}. We derived the analytic solution for $R$ in our previous work \citep{Hsin2022}. The solution for the transition matrix $Z$ can be solved by using Lagrange multipliers, which was derived in \citep{Rabiner1989}. Below, we summarize the analytic solutions for updating each parameter:
\begin{align}
\label{A-sol} A_j  & = \left(\sum_{t=2}^{T} g_t^j \hat{P}_{t,t-1} \right) \left(\sum_{t=2}^{T} g_t^j\hat{P}_{t-1}\right)^{-1}\\ 
\label{Q-sol} \Sigma_j & = \frac{1}{\sum_{t=2}^{T} g_t^j} \left(\sum_{t=2}^{T} g_t^j\hat{P_t}  - A\sum_{t=2}^{T} g_t^j\hat{P}_{t,t-1}^{'}\right)\\ 
B_j & = \left( \sum_{t=1}^{T} g_t^j y_t \hat{x_t}' \right)  \left(\sum_{t=1}^{T} g_t^j \hat{P_t}\right)^{-1}\\
R & = \frac{1}{\sum_{t=1}^{T} \sum_{S_t=j} g_t^j }\left(\sum_{t=1}^{T} \sum_{S_t=j} g_t^j(y_t y_t' - B_j \hat{x_t} y_t')\right)\\
Z_{ij} & = \frac{1}{\sum_{t=2}^{T} g_t^i} \left(\sum_{t=2}^{T} P(S_{t-1} = i, S_t = j | y_{1:T}) \right)
\end{align}
where $i, j = 1, \ldots, M$, 
\begin{align*}
\hat{x_t} & =\text{E}[x_t|y_{1:T}] = \tilde{x}_t,\\ 
\hat{P_t} & = \text{E}[x_t x_t' |y_{1:T}] = \tilde{x}_t \tilde{x}_t' + \tilde{V}_t,\\
\hat{P}_{t,t-1} & = \text{E}[x_t x_{t-1}' |y_{1:T}] = \tilde{x}_t \tilde{x}_{t-1}' + \tilde{V}_{t,t-1}. 
\end{align*}
Note that $\tilde{x}_t, \tilde{V}_t$, and $\tilde{V}_{t,t-1}$ are obtained from the Smoother and Collapse operators as described in \ref{latent_est}. Closed-form solutions for the mean and covariance, $\mu_1$ and $C_1$ for the initial oscillatory state $x_1$ follow the standard derivation for a Gaussian mixture model \citep{Hamilton1990, XuJordan1996}.

In Sections \ref{cnm} and \ref{dim}, as we introduced the structures of $\Sigma_j$ and $A_j$; these matrices are composed of scaled rotation matrices in each $2 \times 2$ off-diagonal block. The analytic solutions for these parameters, as shown in equations \ref{A-sol} and \ref{Q-sol}, do not guarantee this specific rotational structure. We therefore added a step to impose the desired rotational structure in $\Sigma_j$ and $A_j$. Recall that a scaled rotational matrix is a $2 \times 2$ matrix of the form $\gamma \cdot R(\psi)$ , where $\gamma$ is the scaling factor, and $R(\psi)$ is a rotation matrix of the form: 
\begin{equation*}
    R(\psi) = \begin{pmatrix}
        \text{cos}(\psi) & -\text{sin}(\psi) \\
        \text{sin}(\psi) & \text{cos}(\psi).
    \end{pmatrix}
\end{equation*}

\noindent In the M-step of the EM algorithm, we start with the analytic solution for $\Sigma_j$ or $A_j$, depending on the model being estimated, either the Correlated Noise Model or the Directed Influence Model. For both models, our goal is to preserve the rotational structure. We consider the projection space for each off-diagonal block in $\Sigma_j$ or $A_j$ to be the space of scaled rotation matrices. Let the singular value decomposition (SVD) of $A_j^{n_1,n_2}$ be $U_{j}^{n_1n_2}S_{j}^{n_1n_2}V_{j}^{n_1n_2*}$, where $*$ is the conjugate transpose \citep{Beaver2006}. Then the projection to the space of scaled rotation matrices is the product $U_{j}^{n_1n_2}V_{j}^{n_1n_2*}$.  To incorporate scaling, we use the geometric mean of the singular values, $\sqrt{s_{j,1}^{n_1n_2} s_{j,2}^{n_1n_2}}$, where $s_{j,1}^{n_1n_2}$ and $s_{j,2}^{n_1n_2}$ are the singular values on the diagonal of $S_j^{n_1n_2}$. Thus, the scaled rotation matrix becomes $\sqrt{s_{j,1}^{n_1n_2} s_{j,2}^{n_1n_2}} U_{j}^{n_1n_2}V_{j}^{n_1n_2*}$.

The diagonal blocks of the state transition matrix, $A_j^{k}$ in equation \ref{A-matrix} are not purely scaled rotational matrices; instead, each one is a scaled rotational matrix from which a scaling factor times an identity matrix has been subtracted. The scaling factors associated with the identity matrix can be obtained, as they derive from the off-diagonal scaling factors. Consequently, performing SVD on $A_j^{k} + \sum_{n_1=k} \sqrt{s_{j,1}^{n_1n_2}s_{j,2}^{n_1n_2}} \cdot I_{2 \times 2}$ yields $U_{j}^{k}S_{j}^{k}V_{j}^{k*}$. The updated diagonal block then becomes $\sqrt{s_{j,1}^{k}s_{j,2}^{k}} U_{j}^{k}V_{j}^{k*} - \sum_{n_1=k} \sqrt{s_{j,1}^{n_1n_2}s_{j,2}^{n_1n_2}}$, where $s_{j,1}^{k}s_{j,2}^{k}$ are the singular values on the diagonal of $S_j^{k}$. This approach ensures that, with each iteration of the EM algorithm, the updated A matrices maintain a desired rotational structure.

Similarly, we perform the same process on each off-diagonal block of $\Sigma_j$ as outlined in equation \ref{Q-sol} during the M-step to ensure that the rotational structure is maintained. Note that there is no concern about the diagonal blocks in $\Sigma_j$, as they are diagonal matrices that simply represent the covariance of each oscillator.

\subsection{Theoretical Cross-spectrum of the State-Space Model}\label{theory_coh}

Understanding the relationships in the frequency domain between time series is important for unraveling the latent dynamics that govern a system's behavior. Frequency domain methods decompose the signal into sinusoids of specific frequencies, and describe the relative contribution of each of these in explaining the total variability in the signal \citep{Kass2014}. Frequency domain methods are extremely common in the analysis of rhythmic neural data \citep{Greenhouse1987, Logothetis2001, Chaumon2009, AlFahoumAlFraihat2014}. The theoretical cross-spectrum is defined as the Fourier transform of the theoretical cross-covariance function. We derive a parametric solution for the cross-spectrum of our general state-space oscillator model, which allows us to visualize the coordinated behavior of any pair of neural sources. Recall that the latent process model is 
\begin{equation*}
    (x_t|x_{t-1},S_t=j) \sim N(A_j x_{t-1},\Sigma_j)
\end{equation*}
and the observation model is 
\begin{equation*}
    (y_t|x_t,S_t=j) \sim N(B_j x_t,R).
\end{equation*}
Conditioning on $S_t = j$, we suppress subscript $j$ in the following derivation. The latent oscillatory state $x_t$ follows an AR(1) process. The AR($k$) model is given by 
\begin{equation}
    x_t + a_1x_{t-1} + \ldots + a_kx_{t-k} = u_t.
\end{equation}
We set $a_1 = -A$, $a_2, \ldots, a_k = 0$.  
Then, the transfer function matrix of this model \citep{Priestley1981} is given by 
\begin{equation}
    \Gamma_x(\omega) = \left[\alpha(e^{-i\omega})\right]^{-1}, 
\end{equation}
where the matrix polynomial $\alpha(z)$ is defined by $\alpha(z) = \sum_{p=0}^{k}a_p z^p$ and $a_0 = I$. By plugging in $a_1 = -A$, $a_2, \ldots, a_k = 0$, we have 
\begin{equation}
    \Gamma_{x}(\omega) = (I-Ae^{-i\omega})^{-1}.
\end{equation}
Then, following \citep{Priestley1981}, we obtain the spectral matrix of $x_t$ as 
\begin{equation}
    h_x(\omega) = \frac{1}{f_s} \Gamma_{x}(\omega) \Sigma \Gamma_{x}^{*}(\omega),
\end{equation}
where $f_s$ is the sampling frequency, $\omega$ is the frequency of interest, and $*$ denotes the conjugate transpose. 

To compute the spectral matrix of observations $y_t$, we first consider the spectral matrix of $Bx_t$, which can be represented as $\sum_{p=-\infty}^{\infty} \mathbf{g}(p)x_{t-p}$ by setting $g(0) = B$ and $g(p) = 0$ for $p\neq 0$, where $g(p) = \{\mathbf{g}_{ij}(p)\}$ is referred to as the impulse response matrix. Following \citep{Priestley1981}, we derive the transfer function matrix as
\begin{equation}
    \Gamma_{Bx,ij}(\omega) = \sum_{p=-\infty}^{\infty} \mathbf{g}_{ij}(p)e^{-i\omega p} = B_{ij}e^{-i\omega \cdot 0} = B_{ij}.
\end{equation}
That is, the transfer function matrix $\Gamma = B$, then the spectral matrix of $Bx_t$ is given by
\begin{equation}
    h_{Bx}(\omega) = B h_x(\omega) B^T.
\end{equation}
Given the observation equation $y_t = Bx_t + v_t$, where $v_t$ represents independent Gaussian noise, the spectral matrix of the observations $y_t$ can be expressed as the sum of the spectral matrix of $Bx_t$ and the spectral contribution of the noise $v_t$. Hence, the spectral matrix of the observations $y_t$ is
\begin{equation}
    h_y(\omega) = B h_x(\omega) B^T + \frac{1}{f_s}R \quad \text{for all}\,\omega.
\end{equation}

With the spectral matrix of the observations, we can then compute the normalized coherence of the observations as follows:
\begin{equation}
    C_{ij}(\omega) = \frac{h_{ij}(\omega)}{\sqrt{h_{ii}(\omega)h_{jj}(\omega)}},
\end{equation}
where $h_{ij}(\omega)$ is the $ij$-th element of $h_y(\omega)$, for $i, j = 1, \ldots, N$.



\section{Simulation Study}
We conducted a simulation study to demonstrate the application and interpretation of the three model structures introduced in Sections \ref{com}, \ref{cnm}, and \ref{dim}. In section \ref{vis}, we generate example data from small networks generated from each model class and demonstrate how inference from the corresponding model is able to capture the true structure of each network and the times when the networks switch. In section \ref{model-eval}, we simulate data from multiple larger networks under each model structure and compare inferences from both the correctly specified and misspecified models as well as standard time-frequency estimators. This process aims to highlight each model's properties and potential applications, rather than to perform a systematic comparative analysis among these three model structures.

\subsection{Model Visualization}\label{vis}
We simulate data for each model structure described in Sections \ref{com}, \ref{cnm}, and \ref{dim}. For each model, we generate data with four nodes and three different switching states. For illustration purposes, we force two switches to occur only at times 80 seconds and 200 seconds, instead of simulating the switching states directly from the model. Note that the estimation procedures do not know when these switches occur and still attempt to estimate switch times. We utilize the EM algorithm to estimate the model parameters as described in Section \ref{parameter_est}. For each set of simulated data, we fit it to its corresponding model, assuming all the parameters are known except for the one that reflects the network structures.  Specifically, for the COM, the CNM, and the DIM, we fit only $B_j$, $\Sigma_j$, and $A_j$, respectively. The initial parameters, whether $B_j$, $\Sigma_j$, and $A_j$ are set without any specific linkage structure or phase relationship.  

\subsubsection{Common Oscillator Model}
\begin{figure}[htbp]
\centerline{\includegraphics[scale=0.6]{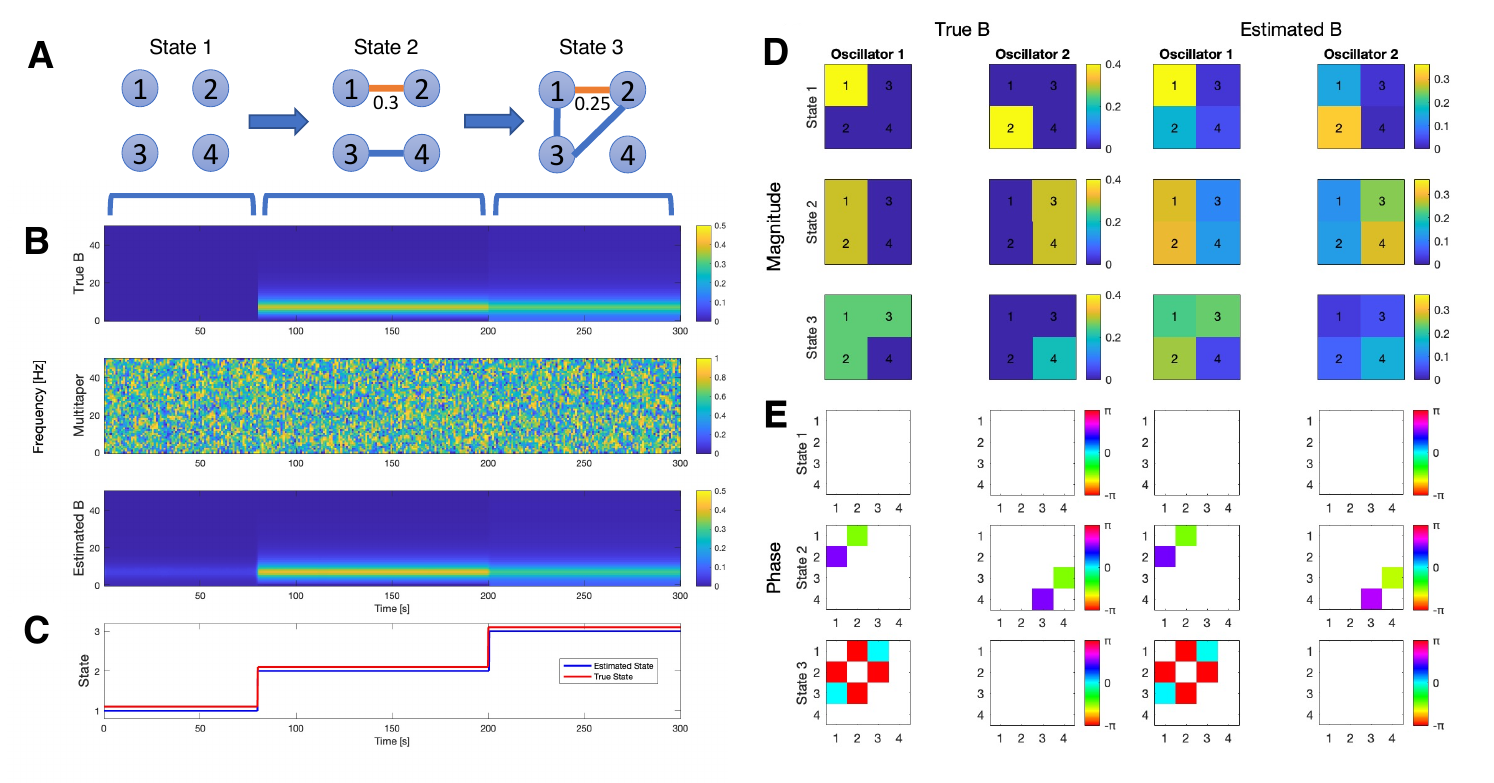}}
\caption{Common Oscillator Model. (A) Diagram of the network structures for states 1, 2, and 3, illustrating how nodes are interconnected in each state. The edge between nodes 1 and 2 is highlighted in orange and annotated with weights, as this pair is analyzed in the coherogram. (B) Coherograms for nodes 1 and 2. The top panel is the theoretical coherogram based on the true network structures and the actual occurrences of switches; the second is generated from a non-parametric multi-taper method estimate of coherence over a sliding window; and the third is the theoretical coherogram based on the estimated parameter, $B_j$, and the estimated switching states. (C) Visualization of the true switching states (red) and the estimated switching states (blue). (D) Visualization of the magnitudes from both the true (left two panels) and estimated (right two panels) observation matrices, $B_j$, highlighting the true and estimated network structure for each state. (E) Visualization of the phase relationship from the true (left two panels) and estimated (right two panels) observation matrices, $B_j$.}
\label{toy-com}
\end{figure}

Starting with the Common Oscillator Model introduced in Section \ref{com}, the network structure is captured by the observation matrix $B_j$, with nodes driven by the same oscillators linked in a network mode. In this simulation, there are four electrodes, two latent states oscillating with a center frequency of 7 Hz, and three switching states, each representing a distinct network structure. The system switches from state 1 to state 2 at 80 seconds and from state 2 to state 3 at 200 seconds. In state 1, node 1 is driven by oscillator 1, and node 2 by oscillator 2, which means these two nodes are not connected. In state 2, nodes 1 and 2 are linked through oscillator 1 with a 90-degree phase difference, while nodes 3 and 4 are connected through oscillator 2, also with a 90-degree phase difference. In state 3, nodes 1, 2, and 3 are driven by the same oscillator, with nodes 1 and 2 being 180 degrees out of phase, nodes 2 and 3 are also 180 degrees out of phase, and nodes 1 and 3 are in phase. Figure \ref{toy-com}A shows the diagram of the three states. If any pair of nodes are connected, they are linked with an edge. We color only the edge between nodes 1 and 2 in orange and label it with weights, indicating linkage strength between nodes 1 and 2, which decreases from 0.3 in state 2 to 0.25 in state 3. This specific coloring is chosen because Figure \ref{toy-com}B focuses on presenting the coherogram for the pair of nodes 1 and 2.

In Figure \ref{toy-com}B, the top panel shows the true theoretical coherogram, accounting for the actual network structure in each state and the occurrences of switches. In states 2 and 3, nodes 1 and 2 are coherent at around 7 Hz, and from state 2 to state 3, the strength of their linkage decreases. The coherogram in the middle panel was generated using a non-parametric multi-taper estimator with a moving window of 1 s and a bandwidth of 2 Hz. We use 1 second windows to have a hope of estimating network switches from second to second, but the resulting estimator is extremely noisy, making it difficult to tell whether the nodes are linked at any frequency. The bottom panel is the theoretical coherogram, constructed based on the estimated parameters of the model and the estimated switching states. The estimates from the COM clearly indicate that nodes 1 and 2 are linked at around 7 Hz in states 2 and 3 and are able to detect the change in linkage strength at time 200 seconds. In Figure \ref{toy-com}C, the true switching states are colored in red, and the estimated switching states are in blue. The estimated states align almost perfectly with the ground truth.

In Figure \ref{toy-com}D, we visualize the magnitude of the parameters that describe the degree to which each oscillator is expressed at each node from equation \ref{B-eq} in Section \ref{com}. The true network structure is shown in the left two panels, where the first and second columns show the extent to which oscillators 1 and 2 are expressed at each node, respectively. The right two panels are the estimated network structures, with two columns representing the two oscillators. Each row represents a single switching state. The estimated model correctly infers that in switching state 1, the first oscillator drives the first node, the second oscillator drives the second node, and these nodes are not linked. Similarly in switching state 2, the model correctly estimates that nodes 1 and 2 are linked through oscillator 1 and that nodes 3 and 4 are linked through oscillator 2. In switching state 3, the model correctly infers that states 1, 2, and 3 are all linked through a common oscillator. In Figure \ref{toy-com}E, we visualize the relative phases of each pair of nodes in a similar manner. For nodes that are driven by the same oscillator, we can compute their phase difference as described in equation \ref{B-eq}. From state 2 to state 3, nodes 1 and 2 are linked with different magnitudes and phases, which can be effectively captured by the COM. By comparing the panels for ground truth with the estimated results, we find that the COM is effective at not only detecting the switches but also estimating the linkage structures and the phase relationships.

In a regime with a low signal-to-noise ratio, the conventional windowed method, the multi-taper coherogram, fails to detect the linkage structure, as shown in the second panel of Figure \ref{toy-com}B. As previously mentioned, the multi-taper coherogram does not assume any specific network structures and allows for any type of link at any time. In contrast, the modeling structures we utilize assume that the system switches between a small number of distinct networks. With switching components, the network structure for each state is estimated over all the time points when that switching state value is correctly estimated. By doing so, we improve the statistical power for estimating network structures and can identify which network is present at any given time, as illustrated in the third panel of Figure \ref{toy-com}B.

\subsubsection{Correlated Noise Model}
Moving to the Correlated Noise Model introduced in Section \ref{cnm}, the network structure is captured by the process noise covariance matrix $\Sigma_j$, with the connectivity of nodes arising from shared noise structure. In this simulation there are four electrodes, and therefore, four latent states oscillating with a central frequency of 7 Hz, where each node is associated with its own oscillator, and three switching states, each representing a unique network structure. Similar to the COM, the system transitions from state 1 to state 2 at 80 seconds, and from state 2 to state 3 at 200 seconds. 

\begin{figure}[htbp]
\centerline{\includegraphics[scale=0.6]{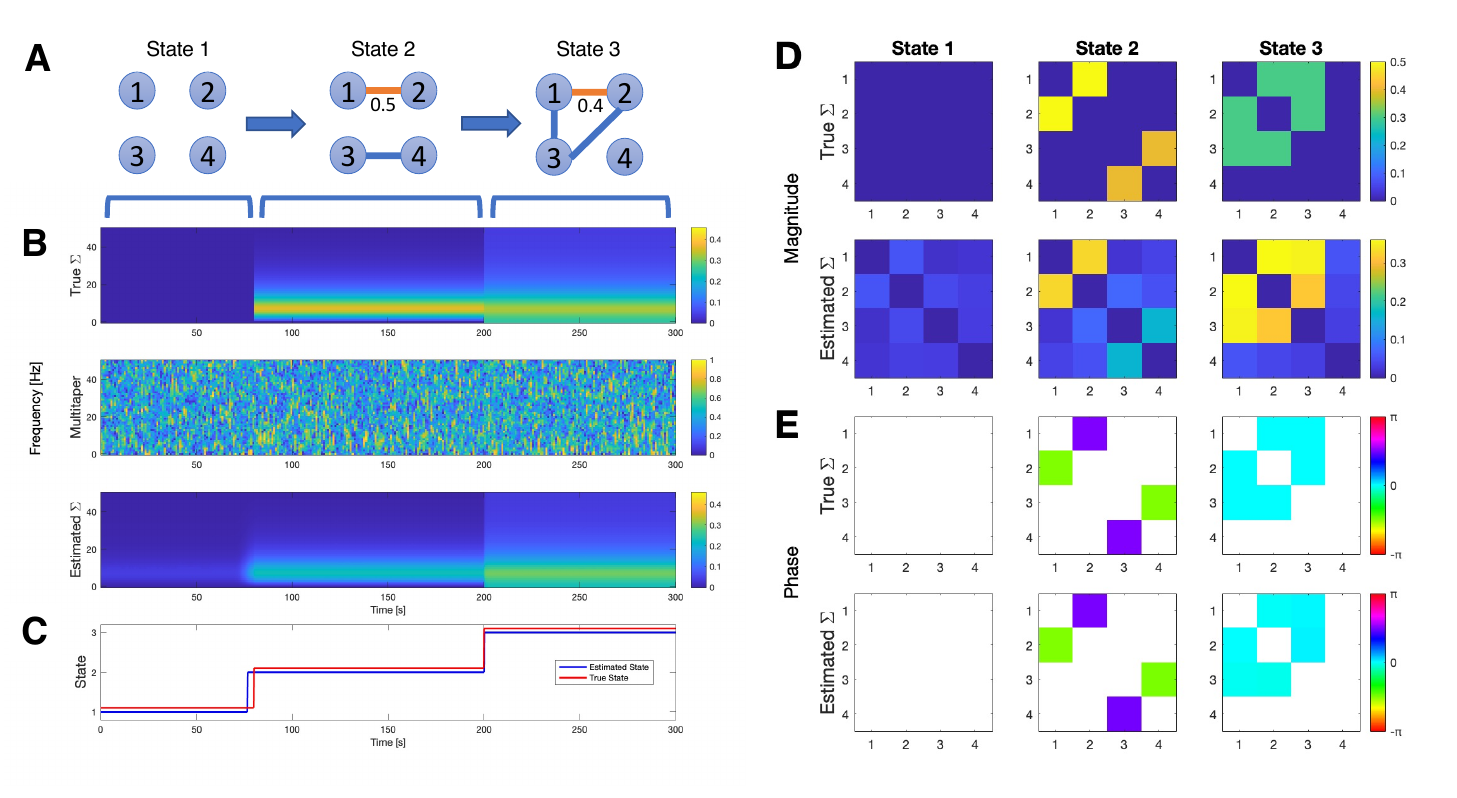}}
\caption{Correlated Noise Model. (A) Diagram of the network structures for states 1, 2, and 3, illustrating how nodes are interconnected in each state. The edge between nodes 1 and 2 is highlighted in orange and annotated with weights, as this pair is analyzed in the coherogram. (B) Coherograms for nodes 1 and 2. The top panel is the theoretical coherogram based on the true network structures and the actual occurrences of switches; the second is generated from a non-parametric multi-taper method estimate
of coherence over a sliding window; and the third is the theoretical coherogram based on the estimated parameter, $\Sigma_j$, and the estimated switching state. (C) Visualization of the true switching states (red) and the estimated switching states (blue) (D) Visualization of linkage magnitudes from the true (top panel) and estimated (bottom panel) noise covariance matrices, $\Sigma_j$, indicating the network structures for each state. (E) Visualization of the phase relationships from the true (top panel) and estimated (bottom panel) noise covariance matrices, $\Sigma_j$.}
\label{toy-cnm}
\end{figure}

In state 1, there are no connections among nodes. In state 2, nodes 1 and 2 are linked with a 90-degree phase difference, and nodes 3 and 4 are connected with a 270-degree phase difference. In state 3, nodes 1, 2, and 3 are interconnected, all in phase. Figure \ref{toy-cnm}A shows the diagram of the three states. Once again, we focus on the pair of nodes 1 and 2, with the corresponding coherograms shown in Figure \ref{toy-cnm}B. The linkage strength for this pair decreases from 0.5 in state 2 to 0.4 in state 3. 

In Figure \ref{toy-cnm}B, we display the true theoretical coherogram (top), the multi-taper coherogram (middle), and the theoretical coherogram from the estimated model (bottom). The multi-taper coherogram is noisy and it is impossible to tell that nodes 1 and 2 are linked at 7 Hz. In contrast, the estimate from the CNM clearly shows that nodes 1 and 2 are linked at around 7 Hz in states 2 and 3, and successfully detects the change in linkage strength at 200 seconds. In Figure \ref{toy-cnm}C, the estimated switching states (blue) nearly perfectly align with the true values of the switching states (red).

In Figure \ref{toy-cnm}D, the true and estimated linkage strengths from equation \ref{Q-matrix} are presented in the top and bottom panels, respectively. From left to right, each column represents a different switching state. The true links are accurately detected by the model estimates. In Figure \ref{toy-cnm}E, we display the true and estimated phase relationships for each pair of nodes, with the true relationships at the top and the estimated ones at the bottom. From state 2 to state 3, nodes 1 and 2 are linked, with their linkage strength decreasing from 0.5 to 0.4 and phase relationship changing from being 90-degree out of phase to in phase. These changes are captured by the CNM.

\subsubsection{Directed Influence Model}
For the Directed Influence Model (DIM) introduced in Section \ref{dim}, we define the network structure via the process transition matrix $A_j$, simulating direct influences among oscillators. Similar to the CNM, each node corresponds to an individual oscillator. In this simulation there are four electrodes, and hence four latent states oscillating with a central frequency of 7 Hz, and three distinct switching states. the system switches from state 1 to state 2 at 80 seconds, and from state 2 to state 3 at 200 seconds. 

\begin{figure}[htbp]
\centerline{\includegraphics[scale=0.6]{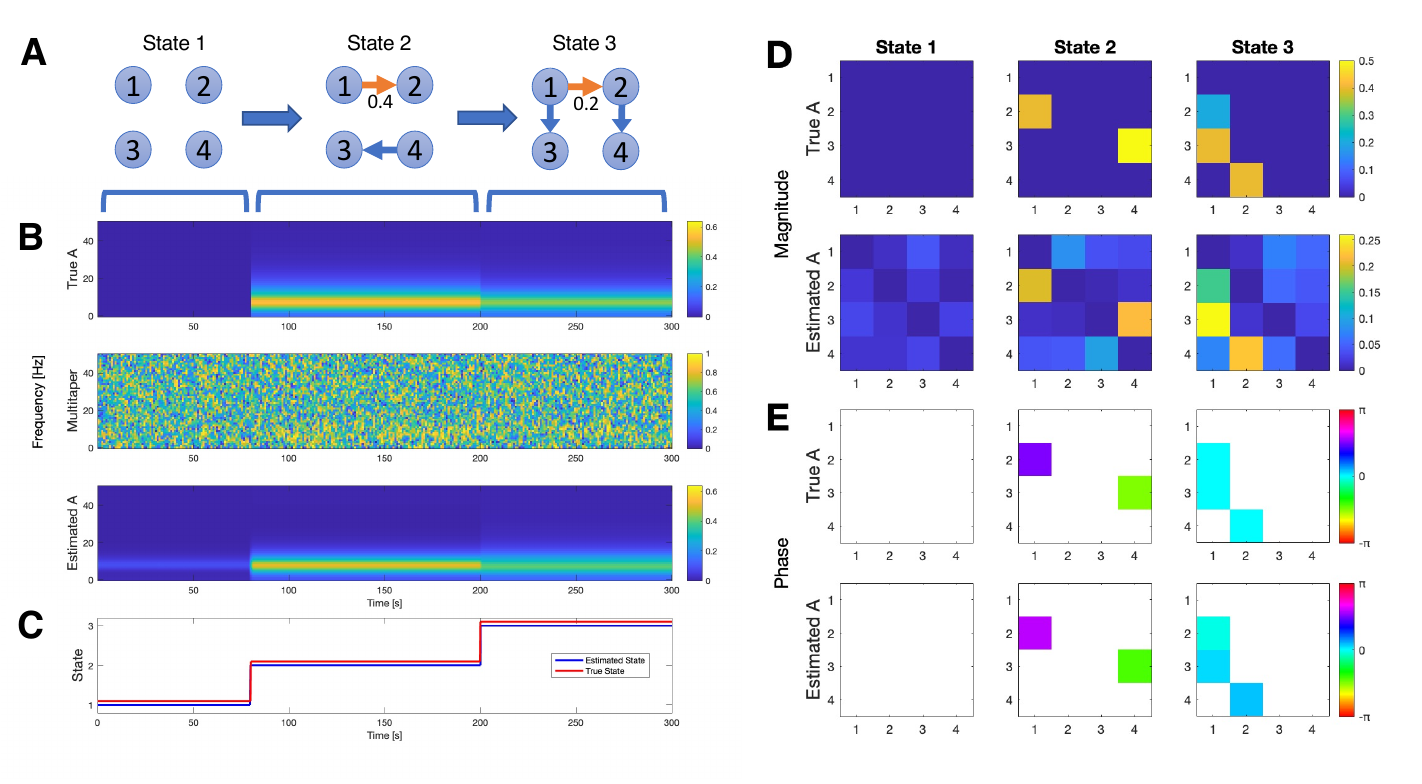}}
\caption{Directed Influence Model. (A) Diagram of the network structures for states 1, 2, and 3, illustrating how nodes are interconnected in each state. The edge between nodes 1 and 2 is highlighted in orange and annotated with weights, as this pair is analyzed in the coherogram. (B) Coherograms for nodes 1 and 2. The top panel is the theoretical coherogram based on the true network structures and the actual occurrences of switches; the second is generated from a non-parametric multi-taper method estimate of coherence over a sliding window; and the third is the theoretical coherogram based on the estimated parameter, $A_j$, and the estimates of switching states. (C) The true switching states are (red) and the estimated switching states (blue). (D) Visualization of magnitude from the true (top panel) and estimated (bottom panel) process transition matrix $A_j$, indicating the network structure for each state. (E) Visualization of the phase relationship from the true (top panel) and estimated (bottom panel) process transition matrix, $A_j$.}
\label{toy-dim}
\end{figure}

In state 1, all nodes remain unconnected. Transitioning to state 2, a directed link is established from node 1 to 2 with a 90-degree phase difference. At the same time, nodes 3 influences 4 with a 270-degree phase difference. In state 3, node 1 influences 2 and 3, and node 2 influences 4, all in phase. The diagram in Figure \ref{toy-dim}A illustrates these networks, with a particular emphasis on the dynamic between nodes 1 and 2, as detailed in the coherograms of Figure \ref{toy-dim}B, where we observe a decrease in linkage strength from 0.4 in state 2 to 0.2 in state 3. As previously mentioned, the DIM is the only model structure we introduce that has directionality. The arrow in the diagram indicates the direction of influence.

Figure \ref{toy-dim}B presents a series of coherograms: the true theoretical, the multi-taper, and the theoretical based on the estimated model. The multi-taper estimate does not indicate a linkage at 7 Hz between nodes 1 and 2 at any time. On the contrary, the estimates from the DIM effectively reveal the connection at 7 Hz between these two nodes in states 2 and 3, capturing the change in linkage strength. The true switching states and the estimated switching states are shown in Figure \ref{toy-dim}C. The estimated states align almost perfectly with the ground truth.

In Figure \ref{toy-dim}D, we visualize both the actual and estimated linkage strengths in the network structures from equation \ref{A-matrix} at the top and bottom, respectively. From left to right, each column represents a different switching state. Figure \ref{toy-dim}E shows the true and estimated phase relationships for each pair of nodes, again with the true relationships at the top and the estimated ones at the bottom. From state 2 to state 3, nodes 1 and 2 are linked, with their linkage magnitude decreasing from 0.4 to 0.2 and phase relationship changing from 90-degree out of phase to in phase. These are captured by the DIM.

The consistent feature in each of these simulations is that the model estimates are able to capture both the network structure in each state and the switches between these networks. The sliding window estimator is able to estimate neither, because short windows do not provide enough data to estimate linkages, and long windows cannot track rapid changes in the network structure. Conversely, by positing that the system switches between a small number of distinct networks, we notably improve our ability to detect the presence and evolution of links between nodes.

\subsection{Model Performance Evaluations}\label{model-eval}
In this subsection, we analyze simulations from larger networks, simulating the switching state $S_t$, the latent oscillatory state $x_t$, and the observation $y_t$ under each model structure, and comparing the estimates using both correctly specified and misspecified models. For each simulation, the number of neural sources is set to ten, the oscillation frequency is centered at 7 Hz, and the sampling frequency is set to 100 Hz. Similar to the previous subsection, we assume all the parameters are known except for the ones that reflect the network structures, and we use the EM algorithm to estimate these parameters. The initial parameters, whether $B_j$, $\Sigma_j$, or $A_j$ are set without any specific linkage structure or phase relationship. 

We examine two evaluation metrics. First, we use the estimated parameters from each model to compute the theoretical cross-spectrum of the fitted model, as described in Section \ref{theory_coh}. We first compute the element-wise difference between the estimated and true cross-spectral matrices at the frequency of interest, and then calculate the root mean square of the magnitudes of all the off-diagonal elements in the difference matrix:
\begin{equation}\label{coh-diff}
    \left\| h_y^{\text{est}}(w) -  h_y^{\text{true}}(w)\right\| = \sqrt{\frac{1}{N(N-1)} \sum_{i \neq j}| h_y^{\text{est}}(w) -  h_y^{\text{true}}(w)|_{ij}^2 },
\end{equation}
where $h_y^{\text{est}}(w)$ is the estimated cross-spectral matrix, $h_y^{\text{true}}(w)$ is the true cross-spectral matrix, $N$ is the number of nodes, and $w$ is the frequency of interest (specifically, $w$ = 7 Hz in simulations). The normalization term is $1/N(N-1)$ because we focus on the off-diagonal terms in the $N \times N$ cross-spectrum matrix. Secondly, we use the fitted parameters to compute the coherence matrix, as described in Section \ref{theory_coh}. We then conduct a hypothesis test on every pair of nodes to determine if they are significantly coherent. This is achieved by fitting an empirical gamma distribution to the off-diagonal elements in the estimated coherence matrix, using a significance level of 0.05. We report the sensitivity and false positive rate. Sensitivity is calculated by computing the proportion of actual coherent links that are correctly identified by our model (true positives) out of all the true coherent links present, according to the true coherence matrix. This metric evaluates how effectively our model identifies coherent pairs of nodes when they are indeed coherent. The corresponding formula for sensitivity is given by:
\begin{equation*}
    \text{Sensitivity} = \frac{\text{True Positives (TP)}}{\text{True Positives (TP)} + \text{False Negatives (FN)}}
\end{equation*}
On the other hand, the false positive rate (FPR) computes the ratio of non-coherent links that are wrongly identified as coherent (false positives) out of all the actual non-coherent pairs, measuring the model's tendency to incorrectly label non-coherent pairs as coherent. The formula for FPR is:
\begin{equation*}
    \text{FPR} = \frac{\text{False Positives (FP)}}{\text{\text{False Positives (FP)}} + \text{True Negatives (TN)}}
\end{equation*}

For assessing the significance of the multi-taper coherence, we perform the F-test with test statistics equal to $(v_0/2-1)|C|^2/(1-|C|^2)$, which is distributed as $F_{2, v_0-2}$ under the null hypothesis of zero population coherence \citep{Hannan1970, JarvisMitra2001}. Here, $v_0/2 - 1$ is the number of tapers and $|C|$ is the magnitude of the multi-taper coherence at the frequency of interest. The derivation of the sample coherence distribution is given in \citep{Hannan1970}, and details of the test statistics can be found in \citep{JarvisMitra2001}. Similarly, we compute the sensitivity and false positive rate for this test.

For example, if the generative model is the COM, we obtain three estimates of the cross-spectral matrix: one from the COM itself, one from the CNM, and one from the DIM. Additionally, we compute the multi-taper cross-spectral matrix in 1 second windows throught the simulation. To compare the cross-spectral estimates from the COM, CNM, and DIM with the true cross-spectral matrix, we compute the norm of the difference matrix, as defined in Equation \ref{coh-diff}, at every time point. We then report the mean and standard deviation of these norm values in Table \ref{table-norm}. Recall that, by conditioning on the switching state $S_t=j$, we can compute the cross-spectral matrix as described in Section \ref{theory_coh}. In our example, with three switching states, this yields three estimated cross-spectral matrices. At every time step, we obtain a single estimated cross-spectral matrix by calculating the weighted sum of cross-spectral matrices from the three states, with the weights based on the probability of the estimated switching state. Similarly, to compare the multi-taper cross-spectral estimates to the true cross-spectral matrix, we compute the norm of the difference matrix at every window and report the mean and standard deviation of these norm values, as shown in the last row in Table \ref{table-norm}. The results of the sensitivity and false positive rate for testing coherence between any pair of nodes in the estimated coherence matrix are presented in Tables \ref{table-sen} and \ref{table-fpr}, respectively. 

\begin{table}[htbp]
\renewcommand{\arraystretch}{1.3}
\caption{Mean and standard deviation (in parentheses) of the element-wise error between the estimated and true cross-spectral matrices (off-diagonal elements). The values are only comparable within each column.}
\label{table-norm}
\begin{center}
\begin{tabular}{|l|c!{\vrule width 2pt}c!{\vrule width 2pt}c|}
\hline
\multirow{2}{*}{Fitted model} & \multicolumn{3}{c|}{Generative model} \\ \cline{2-4}
& DIM & CNM & COM \\ \hline
DIM & 0.0069 (0.0007) & 0.0192 (0.0066) & 0.0030 (0.0008) \\ 
CNM & 0.0093 (0.0013) & 0.0085 (0.0011) & 0.0030 (0.0011) \\ 
COM & 0.0145 (0.0014) & 0.0180 (0.0018) & 0.0015 (0.0003) \\ 
Multi-taper & 0.0221 (0.0021) & 0.0313 (0.0040) & 0.0072 (0.0014) \\ \hline
\end{tabular}
\end{center}
\end{table}

\begin{table}[htbp]
\renewcommand{\arraystretch}{1.3}
\caption{Sensitivity of tests for coherence between pair of nodes with true links, with fractions shown in parentheses.}
\label{table-sen}
\begin{center}
\begin{tabular}{|l|c|c|c|}
\hline
\multirow{2}{*}{Fitted model} & \multicolumn{3}{c|}{Generative model} \\ \cline{2-4}
& DIM & CNM & COM \\ \hline
DIM & 1.000 (40/40)   & 1.000 (54/54)  & 0.917 (22/24) \\ 
CNM & 1.000 (40/40)   & 1.000 (54/54)  & 0.917 (22/24) \\ 
COM & 0.900 (36/40)   & 0.963 (52/54)  & 1.000 (24/24) \\ 
Multi-taper & 0.345 (1404/4066) & 0.364 (1966/5396)& 0.446 (1114/2496)  \\ \hline
\end{tabular}
\end{center}
\end{table}

\begin{table}[htbp]
\renewcommand{\arraystretch}{1.3}
\caption{False positive rate of tests for coherence between pairs of nodes without a link, with fractions shown in parentheses.}
\label{table-fpr}
\begin{center}
\begin{tabular}{|l|c|c|c|}
\hline
\multirow{2}{*}{Fitted model} & \multicolumn{3}{c|}{Generative model} \\ \cline{2-4}
& DIM & CNM & COM \\ \hline
DIM & 0.052 (12/230) & 0.056 (12/216) & 0.057 (14/246) \\ 
CNM & 0.043 (10/230) & 0.046 (10/216) & 0.049 (12/246)) \\ 
COM & 0.052 (12/230) & 0.046 (10/216) & 0.041 (10/246)\\ 
Multi-taper & 0.097 (2230/22934) & 0.129 (2786/21604) & 0.074 (1816/24504) \\ \hline
\end{tabular}
\end{center}
\end{table}

In Table \ref{table-norm}, for data generated by any one of these models—DIM, CNM, or COM — estimates from and of these three models all have smaller error values than those of the multi-taper method, even when the estimated model is misspecified. In Table \ref{table-sen}, we observe a massive improvement in the ability to identify the linkage structures in the data when compared to the multi-taper method, again, even for misspecified models. Note that for the multi-taper method, we perform the coherence test on each window of the tapered segments of the signals. In both tables, it is evident that when the data is fitted with its corresponding generative model, these models successfully capture the structures they are designed to detect and have the smallest error values in each set of the simulated data. Table \ref{table-fpr} shows that the false positive rates for all of the models are well controlled. However, the false positive rates for the multi-taper method are less well controlled. For example, the FPR for multitaper estimates from data generated by the CNM is 0.129, which does not adequately control the FPR to the desired significance level of 0.05. The potential reason for this is that each test is based on a small window, resulting in a lack of accuracy in the approximated F distribution.

As expected, we consistently find that estimation of network structure using the true model that generated the data performs as well or better than estimation from the misspecified models. This suggests that none of these models is superior to the others irrespective of the structure in the data. However, what is more striking is the fact that estimation of the network structure is nearly perfect in these simulations, even when using misspecified models. This highlights the flexibility and general applicability of the models, even in cases where there is minimal prior knowledge about the mechanisms underlying functional connectivity. Each of these model structures takes advantage of the increased statistical power gained by combining information over time. Conversely, if one has specific hypotheses about the mechanisms of functional connectivity, drawing inferences from the corresponding model leads to even better statistical power. 

For each simulated dataset and fitted model, we also report the accuracy of the switching state estimate by calculating the percentage of time steps where the state is correctly estimated, as shown in Table \ref{table-sw}. These models are able to detect the switching states almost perfectly when using the corresponding generative model and still perform well, with accuracy over 95\%, when using a misspecified model. Note that the denominator is 30,000 because we simulate 300 seconds of data with a sampling frequency of 100 Hz, resulting in 30,000 data points.

\begin{table}[htbp]
\renewcommand{\arraystretch}{1.3}
\caption{Fraction of switching states that are correctly identified, with frequencies shown in parentheses.}
\label{table-sw}
\begin{center}
\begin{tabular}{|l|c|c|c|}
\hline
\multirow{2}{*}{Fitted model} & \multicolumn{3}{c|}{Generative model} \\ \cline{2-4}
& DIM & CNM & COM \\ \hline
DIM  & 0.999 (29970/30000) & 0.967 (29001/30000) & 0.999 (29970/30000)\\ 
CNM  & 0.998 (29940/30000) & 0.999 (29977/30000) & 0.999 (29967/30000) \\ 
COM  & 0.997 (29906/30000) & 0.995 (29844/30000) & 0.999 (29968/30000)\\  \hline
\end{tabular}
\end{center}
\end{table}

\begin{figure}[htbp]
\centerline{\includegraphics[scale=0.7]{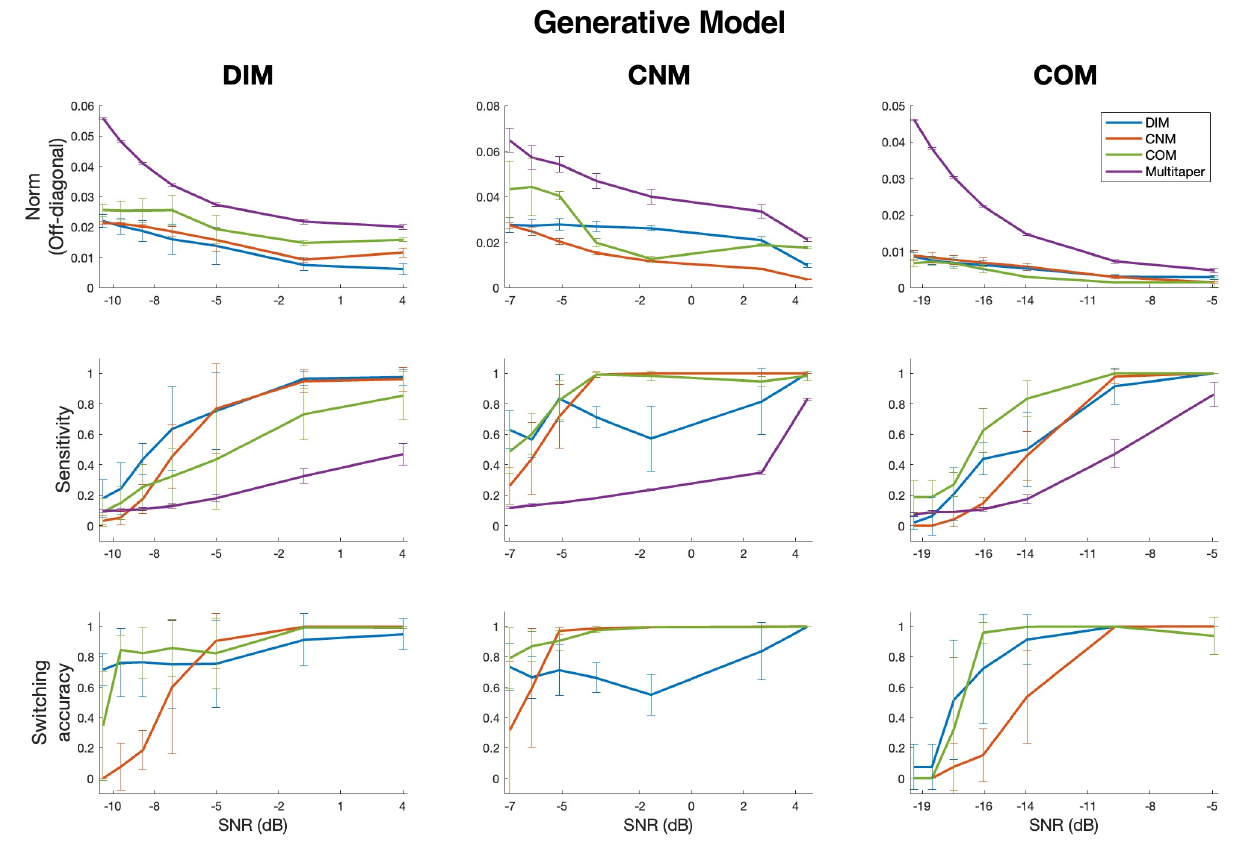}}
\caption{Model evaluation metrics vs the signal to noise ratio (dB) of the simulated data.}
\label{cross-tuning}
\end{figure}

The results from Tables \ref{table-norm}-\ref{table-sw} are in a simulation regime with low signal-to-noise, where tests based on the multi-taper coherogram consisently fail to properly detect the linkage structure in the data. Next, we examined the models' performance by adjusting the signal-to-noise ratio (SNR) and observing how error in the estimate cross-spectrum, sensitivity for detecting links, and switching accuracy are affected. In Figure \ref{cross-tuning}, each column corresponds to a different generative model for the simulated data. The top row shows the error in the estimated cross spectrum when fitting with each model or using the multitaper estimator. As expected, the error values decrease as the SNR increases for every estimation method. We observe that the multi-taper estimator consistently has higher error values than any of the three models. When the fitted model is correctly specified, it consistently has the smallest error values, but the differences between the models tends to be small compared to the error of the multitaper estimator. The middle row shows the sensitivity for detecting true links using each model and the multitaper estimator. While all of the estimators have low sensitivity in our lowest SNR regime, we observe that the sensitivity of the multi-taper method increases much more slowly as the SNR increases than do the model-based estimates. The bottom row shows the accuracy of the estimated switch state for each of the models. Note that the multitaper method does not provide a simple method for estimating which network is expressed at each time. Not surprisingly, we find that switch detection is inaccurate at the lowest SNR regimes and increases rapidly with SNR. Interestingly, we find that for the correctly specified model, switching state estimation starts increasing at lower SNR levels than sensitivity of estimating network structure. this suggests that even when completely accurate inference of the networks is impossible, it may still be possible to detect when changes in network structure occur. 

\section{Discussion}
In this paper, we introduce a broad modeling framework for analyzing dynamical functional connectivity in rhythmic neural data, and investigate three specific model structures, each of which describes different mechanisms of coupling. Broadly, the model framework assumes that a small number of discrete network structures are present at any given time, and maximizes statistical power by estimating network structure over all periods where the same network is determined to be expressed, and determining these periods using the improved network estimates. These approaches allow us to more effectively analyze dynamically coupled oscillators by estimating the underlying functional network structure and identifying changes in their expression. Our purpose in proposing multiple model structures is not to suggest that any one of these is superior to the others, but to demonstrate the inferential value of each and to highlight the improvement in statistical power derived from the broader model framework. 

We believe that the distinct model structures proposed are suited to addressing distinct physiological questions and have the potential to provide different inferential insights. From a physiological perspective, researchers might be interested in specific questions about the mechanisms by which functional oscillatory networks are formed in the brain. Possible explanations include generation from a single source that propagates outward, generation across multiple locations influenced by shared, unobserved inputs, or generation through the direct action of anatomically linked sources. These scenarios might guide them towards the Common Oscillator Model (COM), the Correlated Noise Model (CNM), and the Directed Influence Model (DIM), respectively. From an inferential perspective, researchers might be interested in describing high-dimensional activity within a large connected network using a low-dimensional manifold (i.e., a small set of oscillators). Conversely, researchers might be interested in making inferences about directional or causal links in a network and their role in shaping coordinated activity. In such cases, the choice of a model structure would reflect the inferential goals of the analysis rather than physiological prior knowledge about data generation mechanisms. The choice of which model to apply depends on the specific questions being asked. However, our simulation results suggest that the inferences from these models are robust, even when there is a mismatch between the model used for fitting and the true data generating mechanism.

Another advantage of this modeling framework is its ability to accommodate oscillators at multiple frequencies. While our simulation study focused on networks linked at a single central frequency, it is simple to include oscillators each with its own central frequency to model transitions between networks that are dominated by distinct rhythms. For example, in the analysis of dynamic brain networks during propofol anesthesia, both alpha rhythms and slow waves are though to entrain distinct networks during different stages of unconsciousness \citep{Stephen2020, Mukamel2014, Purdon2013}. Another example is the analysis of signals in the basal ganglia of Parkinson's disease (PD) patients, where distinct networks driven by beta and gamma oscillations have been observed to predict features of motor control tasks \citep{Hirschmann2011, Litvak2011, Swann2016,McGregorNelson2019}. The most commonly reported change is an increase in beta band (13-30 Hz) activity in the motor cortex and basal ganglia when the patient is experiencing motor impairments such as rigidity and slowing of movement (bradykinesia) \citep{LittleBrown2014, Vinding2020}. Conversely, during movement initiation or execution, there is a notable decrease in beta oscillations and an increase in gamma band activity, reflecting the kinetic processes underlying motor control \citep{Brown2003, vanWijk2017}. The multiplicity of frequencies is crucial for modeling and understanding the physiological basis of different brain states. By allowing for the simultaneous representation of these different frequencies, this modeling framework enables us to better interpret the complex patterns of brain activity.

There are a number of ways in which this work may be extended. For the COM, the number of oscillators can be much smaller than the number of neural sources being recorded. In each of our simulations, we fixed this number and used the correct number of oscillators when using the COM to estimate network structure. In prior work, \citep{Hsin2022} we analyzed EEG data during anesthesia induction and return to consciousness using the COM with four oscillators - one for the alpha rhythm and three for the slow waves - to describe the data across 64 electrodes. It is noteworthy that tools have been developed for estimating the number of oscillators from time series data based on this model framework. This includes the oscillator decomposition method \citep{MatsudaKomaki2017b}, which has been employed to extract oscillation components from functional near-infrared spectroscopy (fNIRS) data \citep{Matsuda2022}. The authors highlight that this method is applicable to local field potential (LFP), EEG, magnetoencephalography (MEG), and electrocorticography (ECoG) data as well. In addition, a probabilistic parametric generative model \citep{DasPurdon2022} offers a Bayesian inference procedure for identifying the optimal number of oscillation sources through empirical Bayes model selection. The effectiveness of this approach has been demonstrated on both simulated and real EEG data. If there is no prior knowledge regarding the optimal number of oscillators for use in the COM, the aforementioned approaches can be referred to determine the appropriate number of oscillators.

The broad model framework we proposed was based on the assumption that one network state is expressed at a time, and the system transitions between different network structures nearly instantaneously when a switch occurs. Both the latent oscillatory model and the observation model are conditioned on a single switching state at any given moment. A natural extension would be to model the data as a superposition of multiple states, where all network modes are present at any time, but the degree of expression of each network structure changes over time. In this case, instead of using a latent discrete random variable to describe a switching state, we could use a multi-dimensional continuous random vector to represent expression of each network at each time step. Then, the latent oscillatory state and the observation model would no longer be conditioned on a single state; instead, they would depend on the expression level of all the network modes at the current time. For example, it is reasonable to model different states of motor control in Parkinson's disease as a changing balance between beta- and gamma-driven networks rather than as an instantaneous transition between the two. 

Modeling high dimensional rhythmic neural activity using a state space paradigm, with latent variables that take advantage of known oscillatory dynamics, provides a general and robust framework for understanding shared oscillatory dynamics. This broader analytical framework allows us to capture various functional network structures, each offering insights into the mechanisms and structures underlying coordinated oscillations. These latent process models can be effective in accurately estimating the functional network structure across a wide range of coupling mechanisms and identifying dynamic changes in network expression related to clinical or cognitive context.

\section*{Code Availability}
The code for all the methods described and simulation studies performed in the paper is publicly available through a MATLAB code repository on GitHub at  https://github.com/Stephen-Lab-BU/Switching\_Oscillator\_Networks/

\section*{Acknowledgments} 
This work was supported by the Simons Collaboration on the Global Brain (521921 and NC-GB-CULM-00002730-06). The authors would like to thank Proloy Das for helpful discussions on simulation studies.

\section*{Appendix}
\subsection*{A1: Filter operator}\label{}
In the standard Kalman filter framework, we assume the latent state $x_t$ and the observations $y_t$ satisfy the following linear equations.
\begin{equation}
    \begin{split}
        x_t & = Ax_{t-1} + u_t \\
        y_t & = Bx_t + v_t,
    \end{split}
\end{equation}
where $u_t \sim N(0,\Sigma)$ and $v_t \sim N(0,R)$. Here, we summarize the filter operator
$$ (x_t, V_t, L_t) = \text{Filter}(x_{t-1}, V_{t-1}, y_t, A, \Sigma, B, R). $$
At each time step $t$, it has two steps: (1) prediction of the state and its error covariance estimates and (2) update these estimates with new measurement data. The computation is based on a recursive approach for the incorporation of the new observation $y_t$. The steps are as follows. We first compute the predicted state $x_{t|t-1}^{-}$ and its error covariance matrix $V_{t|t-1}^{-}$ as follows:
\begin{equation}
\begin{split} 
    x_{t|t-1}^{-} & = A x_{t-1} \\
    V_{t|t-1}^{-} & = A V_{t-1} A^{'} + \Sigma
 \end{split}
\end{equation}
We then compute the innovation $e_t$ (the state error in the prediction), the Kalman gain matrix $K_t$, the updated state estimate $\hat{x}_t$, and the updated state error covariance matrix $V_t$.
\begin{equation*}
\begin{split} 
    e_t & = y_t - B x_{t|t-1}^{-} \\
    K_t & = V_{t|t-1}^{-} B^{'} (B V_{t|t-1}^{-} B^{'} + R)^{-1}\\
    \hat{x}_t & = x_{t|t-1}^{-} + K_t e_t\\
    V_t &= (I-K_t B) V_{t|t-1}^{-}
 \end{split}
\end{equation*}
In the prediction step, the equations use the system's model to forecast the next state and its error covariance. The update step adjusts these predictions based on the new measurement $y_t$, with the Kalman Gain $K_t$ determining the weight of the adjustment. This process iteratively refines the state estimated by balancing between the predictions and the measurements. The likelihood $L_t$ of observing the observation $y_t$ can be computed by $$L_t = N(e_t; 0, B V_{t|t-1}^{-} B^{'} + R).$$

\subsection*{A2: Smoother operator}

We denote the smoother operator:
$$ (\tilde{x}_t, \tilde{V}_t, \tilde{V}_{t, t-1}) = \text{Smoother}(\tilde{x}_{t+1}, \tilde{V}_{t+1}, x_t, V_t, A, \Sigma). $$ To apply the smoother algorithm, we start with the final estimates from the Kalman filter for the state $x_T$ and covaraince $V_T$ at the last time step $T$. Then, moving backward from time step $t=T$ to $t=1$, for the prediction step, we can pass the estimates in from the filtering stage. Then, we compute the smoother gain matrix $G_t$, the updated state estimate $\tilde{x}_t$, the updated state covariance estimate $\tilde{V}_t$, and the one-step covariance estimate $\tilde{V}_{t, t-1}$ as follows:
\begin{equation*}
\begin{split} 
    G_t & = V_t A^{'} V_{t+1}^{-1} \\
    \tilde{x}_t & = \hat{x}_t + G_t(\tilde{x}_{t+1}-\hat{x}_{t+1})\\
    \tilde{V}_t & = V_t + G_t(\tilde{V}_{t+1}-V_{t+1})G_t^{'} \\
    \tilde{V}_{t, t-1} & = V_t G_{t-1}^{'} + G_t(\tilde{V}_{t+1, t} - Av_t) G_{t-1}^{'}
 \end{split}
\end{equation*}
The computation of the one-step covariance estimates shown above is an alternative way that does not required the corresponding filtered terms \citep{ShumwayStoffer1991}, where the initial condition is $$\tilde{V}_{T, T-1} = (I-K_T B) A V_{T-1}.$$ The smoother gain matrix $G_t$ is used to adjust the state estimate $\hat{x}_t$ with information from the future state $\tilde{x}_{t+1}$. The estimates $\tilde{x}_t$, $\tilde{V}_t$ and $\tilde{V}_{t, t-1}$ represent the smoothed estimates by taking into account all observations up to time $T$. This backward pass through all the data allows the smoother algorithm to provide a more accurate estimate of the state at each time step by incorporating information from the entire dataset, rather than just the information available up to that point in time as in the filtering stage.

\subsection*{A3: Collapse and Cross-Collapse operators}\label{collapse}

The Collapse and Cross-Collapse operators are used to approximate a mixture of Gaussians as a single Gaussian by moment matching. Given the conditional mean for each state is $x_t^j = \text{E}(x_t|S_t = j)$, its covariance $V_t^j = \text{Cov}(x_t|S_t = j)$, and the mixing coefficient $g_t^j = P(S_t = j)$ as the inputs of the operator, we can compute the unconditional moments as follows. For the collapse operator $$(x_t, V_t) = \text{Collapse}(x_t^j, V_t^j, g_t^j),$$ we compute 
\begin{align*}
    x_t & = \sum_{j} g_t^j x_t^j \\
    V_t & = \sum_{j} g_t^j (V_t^j - (x_t^j - x_t)(x_t^j - x_t)^{'}).
\end{align*}

\noindent For the cross-collapse operator $$(x_{t+1}, x_t, V_{t+1,t}) = \text{Cross-Collapse}(x_{t+1}^j, x_t^j, V_{t+1,t}^j, g_t^j),$$
we compute
\begin{align*}
    x_{t+1} & = \sum_{j} g_{t+1}^j x_{t+1}^j \\
    x_t & = \sum_{j} g_t^j x_t^j \\
    V_{t+1,t} & = \sum_{j} g_t^j (V_{t+1,t}^j - (x_{t+1}^j - x_{t+1})(x_t^j - x_t)^{'}).
\end{align*}

\noindent It has been demonstrated that a Gaussian with these moments is the nearest possible Gaussian approximation to the original mixture distribution \citep{Lauritzen1996, Murphy1998}.

\bibliographystyle{APA}

\end{document}